\documentclass[usenatbib,useAMS]{mn2e}
\usepackage{times}
\usepackage{epsfig}

\newcommand{\tens}[1]{\mathbfss{#1}}
\newcommand{\tenscp}[1]{\mathsf{#1}}
\renewcommand{\vec}[1]{\bmath{#1}}

\begin{document}
\title{Theory and practice of microlensing lightcurves around fold singularities}
\author[M. Dominik]{M.~Dominik \\
University of St Andrews, School of Physics \& Astronomy, North Haugh,
St Andrews, KY16 9SS, United Kingdom}
\maketitle

\begin{abstract}
Among all galactic microlensing events, those
involving a passage of the observed source star over the caustic created
by a binary lens are particularly useful in providing 
information about stellar atmospheres, the dynamics of stellar populations in our own and
neighbouring galaxies, and the statistical properties of stellar and sub-stellar binaries. 
This paper presents a comprehensive guide for modelling and interpreting the lightcurves 
obtained in events involving fold-caustic passages.
A new general, consistent, and optimal choice of parameters provides 
a deep understanding of the
involved features, avoids numerical difficulties and
minimizes correlations between model parameters. 
While the photometric data of a microlensing event around a caustic passage itself
do not provide constraints on the characteristics of the underlying binary lens
and do not allow predictions of the behaviour of other regions of the lightcurve, 
vital constraints can be obtained in an efficient way if these are combined with a few
simple characteristics of data outside the caustic passages. A corresponding algorithm 
containing some improvements over an earlier approach which  
takes into account multi-site observations 
is presented and discussed in detail together with the arising parameter constraints
paying special attention to the role of source and background fluxes.
\end{abstract}


\section{Introduction}

The majority of the galactic microlensing events that have been detected or are 
currently detected
are compatible with both the observed source star and the compact massive lens being approximated 
by point-like objects. However, about 2 to 3~per cent of the events 
involve peaks lasting from a
few hours to a few days whose shape is characteristic 
for the source crossing a  
fold-caustic line
which is created by a binary (or multiple) lens object.
These fold-caustic passage events provide valuable information about both the source star and the
lens object not being extractable from 'ordinary' microlensing events.

Microlensing events compatible with a point-like source at distance $D_\rmn{S}$ and 
a point-like lens of mass $M$ at distance $D_\rmn{L}$ involve only one parameter
that is related to physical properties of lens or source and
carries a dimension, namely the time-scale of motion $t_\rmn{E} = \theta_\rmn{E}/\mu$, where
$\mu$ denotes the proper motion of the source relative to the lens and $\theta_\rmn{E}$ is
the angular Einstein radius, given by
\begin{equation}
\theta_\rmn{E} = \sqrt{\frac{4GM}{c^2}\,\frac{D_\rmn{S}-D_\rmn{L}}{D_\rmn{S}\,D_\rmn{L}}}\,.
\label{eq:angEinstein}
\end{equation}
The time-scale of motion $t_\rmn{E}$ is therefore a convolution of the
lens mass $M$, the lens distance $D_\rmn{L}$, 
and the proper motion $\mu$, which cannot be measured individually, so that
the power of ordinary microlensing events for the determination
of the mass function and phase-space distribution of stellar populations in our own or 
neighbouring
galaxies that have caused these events is severely limited \citep{MP:massdet}.
In contrast to this, the observed duration of a fold-caustic passage $t_\star^\perp$ 
provides a second time-scale which
is related to the angular size of the source star $\theta_\star$.
An estimate for $\theta_\star$  can be obtained from the stellar flux
and spectral type,
so that $t_\star^\perp$ provides a lower limit to the proper motion $\mu$ by 
yielding its component $\mu^\perp = \theta_\star/t_\star^\perp$
perpendicular to the caustic, and $\mu = \mu^\perp/(\sin \phi)$ itself follows with a determination
of the crossing angle $\phi$ from modelling the full observed lightcurve.
The measurement of the relative proper motion between lens and source for a
microlensing event toward the Small Magellanic Cloud (SMC) 
\citep{EROS:SMC,PLANET:SMC,PLANET:sol,joint,MACHO:SMC,MPS:SMC,OGLE:SMC}
indicated that the lens in located in the SMC rather than in the Galaxy, 
giving support to the
hypothesis that the majority of microlensing events toward the SMC are due to self-lensing among its stellar
populations \citep[e.g.][]{Stef:SMC}. 

In addition to providing a measurement of the proper motion $\mu$, lightcurves around
caustic passages are particularly sensitive to effects caused by the parallactic motion of the
Earth around the Sun or by the orbital motion of the binary lens
\citep{HW95,GA99}. Observations of these effects
and the determination of the corresponding parameters \citep[e.g.][]{Do:Rotate}
together with the determination of the
proper motion $\mu$ break the degeneracy between the lens mass $M$ and its distance $D_\rmn{L}$
yielding measurements of the individual quantities 
(up to possible discrete ambiguities).
One such measurement has been reported by \citet{PLANET:EB5mass}.

The characteristic and distinctive shape of the lightcurve around caustic passages 
allows to obtain stronger constraints on the parameters of the
binary lens than those resulting from binary lens events that show only weak deviations compared
to single lenses. Therefore fold-caustic passage events provide the main source of information about
the statistical distributions of the properties of stellar and sub-stellar binaries such 
as the total mass, mass ratio, semimajor axis, or orbital period.  

The most valuable results arising from caustic-passage microlensing events so far have been on 
the study of stellar atmospheres. The strong differential magnification that arises as the 
source passes over the caustic allows to resolve surfaces of source stars at Galactic distances
providing a powerful and unique technique to study stellar atmospheres of
a variety of different types of stars thereby probing existing theoretical models 
\citep{SchneiWei:QSO,Rhie:ld,GG:detld}.
Measurements of limb-darkening coefficients from binary lens microlensing events
have been published for three K-giants
\citep{PLANET:M28,PLANET:M41,PLANET:EB5}, one G/K-subgiant \citep{PLANET:O23}
and a Solar-like star \citep{Abe} in the Galactic bulge 
as well as for an A-dwarf in the SMC \citep{joint}, where the source sweeps over
a cusp for two of these events. 
In addition to a dense photometric coverage during the course of a caustic passage, 
the observation of a temporal sequence of spectra
provides a probe of the spatial distribution of chemical elements 
\citep{GG:detld} using the observed
variation of related spectral lines
as indicators. For a galactic bulge microlensing event in 2000, several 
low-resolution spectra have been taken with the FORS1 spectrograph at the VLT (ESO Paranal, Chile)
 showing a significant variation of the equivalent width of the
H$\alpha$-line by \citet{PLANET:EB5spect}, while
two high-resolution spectra have been taken at the Keck telescope (Mauna Kea, Hawaii, USA)
by \citet{Keck:EB5spect}, and both measurements have later been 
discussed by \citet{Afonso:EB5}. In 2002, 
a dense sequence of
high-resolution spectra has been obtained for another event with 
the UVES spectrograph at the VLT
where significant equivalent-width variations in the most prominent lines H$\alpha$, H$\beta$,
and \mbox{Ca\,\sc{ii}} have been detected
\citep{PLANET:OB69letter,PLANET:OB69}. 

In addition to the events already discussed in the literature,
a dense high-precision photometric coverage of caustic-passage regions of
several further microlensing events 
allowing the determination of limb-darkening coefficients
has
been obtained by PLANET\footnote{\tt http://planet.iap.fr} 
\citep{PLANET:first,PLANET:EGS} during recent years, and additional
data on caustic-passage events have also been collected by other microlensing collaborations,
namely EROS\footnote{\tt http://eros.in2p3.fr}
 \citep{EROS:general}, 
MACHO\footnote{\tt http://wwwMACHO.mcmaster.ca} \citep{MACHO:binaries,MACHO:generalBulge,MACHO:generalLMC}, 
OGLE\footnote{\tt http://www.astrouw.edu.pl/\~{}ftp/ogle} \citep*{OGLE:general1,OGLE:general2}, 
MOA\footnote{\tt http://www3.vuw.ac.nz/scps/moa} \citep{MOA:general}, 
MPS\footnote{\tt http://bustard.phys.nd.edu/MPS} \citep{MPS:SMC},
and MicroFUN\footnote{{\tt http://www-astronomy.mps.ohio-state.edu/}\\\hspace*{2em} {\tt\~{}microfun}} \citep{MicroFUN}.
 With currently $\sim\,$500 microlensing alerts per year issued by 
OGLE-III\footnote{{\tt http://sirius.astrouw.edu.pl/\~{}ogle/ogle3/}\\
\hspace*{2em} {\tt ews/ews.html}} and
$\sim\,$60 alerts per year issued by MOA\footnote{{\tt http://www.roe.ac.uk/\~{}iab/alert/alert.html}},
about 10--15 caustic-passage events per year 
can be expected for which limb-darkening coefficients of the source star
can be obtained by
a dense high-precision photometric coverage, while, 
depending on allocated
resources, spectroscopic measurements should be feasible 
for a few events per year. 

The aim of this paper is to provide a comprehensive guide for modelling and interpreting
microlensing lightcurves near fold singularities which are produced by extended sources and
binary lenses, and for finding all suitable models for microlensing events involving
fold-caustic passages. In addition, the derivation of physical properties
of the lens and the source resulting from the modelling of such events is discussed.
This paper outlines the underlying theory and sets the notation for
upcoming analyses of observed data. A similar paper about cusp 
singularities including a comparison with the relations for
folds as discussed in this paper is under preparation.

For lightcurves near a fold-caustic passage, a new consistent, general, and optimal set
of parameters which are directly related to observable features of the observed
lightcurve is introduced. This choice not only leads to a better understanding
of the features of the lightcurve by allowing an obvious interpretation in terms
of corresponding properties of lens or source, but also minimizes
correlations between the parameters. In addition, singularities in parameter space
likely to cause numerical problems, e.g.~when the source size tends to zero or the
background flux assumes some special values (in particular for flux values
arising from image-subtraction), are avoided.

Finding all suitable model parameters corresponding to lightcurves consistent with the
observed data of a complex microlensing event is a non-trivial task due to the
large number of parameters and intrinsic degeneracies and ambiguities between
the physical properties of the source and the lens system 
\citep{MaoDiStef,DoHi2,DiStefPerna,Do:Ambig,Do99:CR,PLANET:sol}. 
Since the lightcurve of sources in the vicinity of the fold is determined by
local properties related to the fold singularity, limb-darkening measurements do not
require the assessment of models for the complete lightcurve along with arising
ambiguities and degeneracies. However, a model of the full lightcurve is needed
for determining the proper motion $\mu$,
the time-scale of motion $t_\rmn{E}$ that carries information about lens mass
$M$ and distance $D_\rmn{L}$, and
the mass ratio $q$ and the separation parameter $d = \delta/\theta_\rmn{E}$ 
of the binary lens (where $\delta$ is the instantaneous
angular separation between its components). 
Moreover, a dense coverage of the lightcurve during the
caustic passage 
and in particular target-of-opportunity observations of spectra yielding
a powerful test of the atmosphere of the source star require some
prior arrangements pushing the need for a prediction of caustic passages.
With the lightcurve near a caustic passage being determined by local properties only,
a prediction is possible with data taken on the rise to a caustic exit, while
the data taken over previous caustic passages itself do not provide any information,
whereas the combination with previous data outside caustic passages can yield some
constraints \citep{PLANET:sol,JM:2ndcc}.

In general, the correlations between model parameters are minimized
by choosing parameters that
correspond to observable characteristics of the observed lightcurve.
For binary lens caustic-passage microlensing events, the caustic passages themselves
show some characteristic features. Restricting the full parameter space by modelling
a caustic passage and performing a search in the lower-dimensional subspace holding
additional parameters necessary to describe the full lightcurve therefore constitutes
an efficient method for modelling such events \citep{PLANET:sol}.
In this paper, an improved variant of this approach 
taking into account multi-site and/or multi-band observations is described and discussed
in detail, and the relations between the fold-caustic model parameters, the properties of
the binary lens, and observable characteristics of the full lightcurve such as
the time-scale of motion or the source and background fluxes are investigated.

In Sect.~\ref{sec:lensmapping}, the basic theory of gravitational lensing is reviewed setting
the basic notation used in this paper, 
while Sect.~\ref{sec:foldsing} gives a detailed discussion of the fold singularity and its properties.
Sect.~\ref{sec:magstar} investigates the magnification of source stars in the vicinity of
fold caustics and Sect.~\ref{sec:magprofiles} discusses the specific implications arising from
limb darkening. Sect.~\ref{sec:parametrization} deals with lightcurves of source stars in a region
close to a caustic passage within a microlensing event and presents an adequate parametrization
of such parts of lightcurves to be used for modelling observed data where the model parameters
directly correspond to observable properties of the lightcurve and can be easily understood. 
The determination of a complete set of parameters characterizing the full lightcurve by making
use of the parameters locally characterizing the caustic region 
forms the content of Sect.~\ref{sec:fullmodel},
where two versions of the basic algorithm (differing in the consideration of the
temporal variation of the magnification of non-critical images)
are discussed, followed by discussions of constraints arising
from non-negative background flux and from simple characteristics of the data outside the
caustic-passage region such as the baseline flux or the time-scale of motion.  
In Sect.~\ref{sec:predpow}, 
the predictive power of the data in the caustic passage region and its combination with 
characteristics of data outside it with respect to the determination of the characteristics
of the binary lens and the perspective for predicting future caustic passages is discussed.
Sect.~\ref{sec:summary} finally provides a summary of the paper and its results. 
A table of the used symbols has been attached at the end of the paper.

\section{The lens mapping}
\label{sec:lensmapping}

Gravitational lensing is understood as
the bending of light emitted by a source
caused by the gravitational field of
intervening matter. Due to this effect, a source located at a position
${\vec y}$ on the sky (in arbitrarily scaled coordinates) will
be observed at one or more image positions ${\vec x}^{(i)}$.
Moreover, the luminosity of the observed images differs from the
intrinsic luminosity of the source object.
In general, a lens mapping can be described by the Fermat-Potential
\begin{equation}
\Phi({\vec x}, {\vec y}) = \frac{1}{2}\,({\vec x} - {\vec y})^2 - \psi({\vec x})
\label{eq:potential}
\end{equation}
\citep{Schnei:potential} corresponding to the arrival time of (hypothetical) 
light rays at the observer.
Fermat's principle determines the actual light rays to satisfy
\begin{equation}
{\vec \nabla}_{{\vec x}} \Phi = 0\,,
\label{eq:raytrace}
\end{equation}
which relates 
source and image positions by
the lens equation
\begin{equation}
{\vec y} = {\vec x} - \vec \alpha({\vec x})\,,
\label{eq:lenseq}
\end{equation}
where
\begin{equation}
\vec\alpha({\vec x}) = {\vec \nabla}_{{\vec x}} \psi({\vec x})\,.
\end{equation}
Let indices to $\Phi$ denote its $n$-th partial derivatives with respect
to $x_{i_1} \ldots x_{i_n}$, $i_k$ being the coordinate index of the
$k$-th derivative, i.e.\ 
\begin{equation}
\Phi_{i_1,\ldots,i_n} = \frac{\partial}{\partial x_{i_n}} \ldots
\frac{\partial}{\partial x_{i_1}}\,\Phi\,,
\label{eq:phiabbr}
\end{equation}
and let $\psi_{i_1,\ldots,i_n}$ denote the corresponding derivatives of $\psi$.
For (locally) continuous derivatives, the order of derivation and therefore the order
of indices to $\Phi$ or $\psi$  is irrelevant
(Schwarz's theorem).
For the second derivatives of $\Phi$, one obtains the relation
\begin{equation}
\Phi_{ij}(\vec x) = \delta_{ij} - \psi_{ij}(\vec x) = \frac{\partial y_i}{\partial x_j}\,,
\label{eq:Phipsi}
\end{equation}
where $\delta_{ij}$ denotes the Kronecker symbol
\begin{equation}
\delta_{ij} = \left\{\begin{array}{lcl}0 & \mbox{for} & i \neq j \\ 
 1 & \mbox{for} & i = j \end{array}\right.\,.
\end{equation}
Eq.~(\ref{eq:Phipsi}) implies that all derivates 
of $\vec y$ with respect to $x_{i_k}$ can be expressed by means of
the Fermat potential $\Phi$ and its derivatives. 
In particular, the components of the
Jacobian matrix $\tens{J}$ read 
\begin{equation}
\tenscp{J}_{ij} = \frac{\partial y_i}{\partial x_j} = 
\Phi_{ij}\,, 
\end{equation}
and its determinant becomes 
\begin{equation}
\det \tens{J} = \Phi_{11}\,\Phi_{22} - (\Phi_{12})^2\,.
\end{equation}

The Jacobian determinant yields the (signed) magnification of an image at ${\vec x}^{(i)}$ as
\begin{equation}
\mu({\vec x}^{(i)}) = \frac{1}{\det \tens{J}({\vec x}^{(i)})} \,, 
\end{equation}
and the total magnification $A$ of a source 
at ${\vec y}$ is obtained by summing over the absolute values
of the individual magnifications of its $N$ images, i.e.
\begin{equation}
A({\vec y}) = \sum\limits_{i=1}^{N} |\mu({\vec x}^{(i)})|\,,
\end{equation}
where ${\vec x}^{(i)}$ and ${\vec y}$ satisfy the lens equation.

The lens mapping becomes singular at 
points ${\vec x}^{(\rmn{crit})}$ in image space for which the Jacobian determinant vanishes
($\det \tens{J}({\vec x}^{(\rmn{crit})}) = 0$) which themselves are called    
critical points. 
The set of critical points forms critical curves which are mapped onto the caustics
by the lens equation. The number of images associated with a given source position
changes (by multiples of two) 
with the source position if and only if the source crosses a caustic. 
Therefore, the caustics divide the two-dimensional space of source positions into
regions with a fixed number of images, where a source is defined to be 'inside'
a caustic if it belongs to the region with the larger number of images, and 
'outside' the caustic otherwise.
The singularities of the lens mapping can be categorized into different types showing 
their own specific characteristics.

\section{The fold singularity}
\label{sec:foldsing}

The lowest-order singularity is the fold, for which
the following conditions need to be fulfilled
\citep{SchneiWei:cusp}:
\begin{itemize}
\item exactly one of the eigenvalues of the Jacobian matrix is zero,
\item the tangent vector of the critical curve is not an eigenvector of the Jacobian 
matrix
belonging to eigenvalue zero.
\end{itemize}
For critical points, the eigenvalues of $\tens{J}$ are given by
\begin{equation}
\lambda^{(1)} = \Phi_{11} + \Phi_{22} = \mbox{Tr}\,{\tens J}\,, \quad \lambda^{(2)} = 0\,.
\end{equation}
The first condition for a fold then implies that
$\Phi_{11} + \Phi_{22} \neq 0$.
Normalized eigenvectors that correspond to these eigenvalues can be written
in the form
\begin{equation}
\vec e^{(1)} = \left(\begin{array}{c}
\cos \theta \\ \sin \theta 
\end{array}\right)\,, \quad
\vec e^{(2)} = \left(\begin{array}{c}
-\sin \theta \\ \cos \theta 
\end{array}\right)\,,
\label{eq:eigenvec}
\end{equation}
where
\begin{eqnarray}
\sin \theta & = & \varepsilon(\Phi_{12}\,\Phi_{22})\,\sqrt{\left|\frac{\Phi_{22}}{\Phi_{11} + \Phi_{22}}\right|}\,, \nonumber \\
\cos \theta & = &\sqrt{\left|\frac{\Phi_{11}}{\Phi_{11} + \Phi_{22}}\right|}\,,
\label{eq:sincosexpl}
\end{eqnarray}
with
\begin{equation}
\varepsilon(z) = \left\{\begin{array}{ccl} 1 & \mbox{for} &
z \geq 0 \\ 
-1 & \mbox{for} & z < 0 
\end{array}\right.\,. 
\label{eq:defeps}
\end{equation}
The vanishing of the Jacobian determinant yields 
$\Phi_{11} \Phi_{22} = (\Phi_{12})^2 \geq 0$, so that
$|\Phi_{11} + \Phi_{22}| = |\Phi_{11}| + |\Phi_{22}|$. 

Let ${\vec x}_\rmn{f}$ denote a point on the critical curve
and ${\vec y}_\rmn{f} =
{\vec x}_\rmn{f} - \vec \alpha({\vec x}_\rmn{f})$ denote the
corresponding source coordinate on the fold caustic. 
The eigenvectors of $\tens{J}$ given by Eq.~(\ref{eq:eigenvec}) then
constitute a right-handed basis of a coordinate system defined by
\begin{eqnarray}
{\vec x}(t) & = & {\vec x}_\rmn{f}+ {\bmath{\mathcal R}}\,{\vec x}'(t)\,, 
\label{eq:xtrafo}\\
{\vec y}(t) & = & {\vec y}_\rmn{f}+ {\bmath{\mathcal R}}\,{\vec y}'(t)\,,
\label{eq:ytrafo}
\end{eqnarray}
where
\begin{equation}
{\bmath{\mathcal R}} = \left(\begin{array}{cc}
\cos \theta & -\sin \theta \\ \sin \theta & \cos \theta
\end{array}\right)\,,
\label{eq:rotmat}
\end{equation}
so that $\theta$ is the orientation of the $(x'_1,x'_2)$- and $(y'_1,y'_2)$-
coordinates with respect to the 
$(x_1,x_2)$- and $(y_1,y_2)$-coordinates, respectively.
Since every multiple of an eigenvector is an eigenvector itself, 
$-(\vec e^{(1)},\vec e^{(2)})$ also constitutes a right-handed basis
being related to $(\vec e^{(1)},\vec e^{(2)})$ by
a $180\degr$-rotation, equivalent to replacing $\theta$ by $\upi + \theta$.
Eq.~(\ref{eq:xtrafo}) implies that
\begin{equation}
{{\mathcal R}}_{ij} = \frac{\partial x_i}{\partial x'_j}\,.
\label{eq:rder}
\end{equation}

The Jacobian matrices $\tens{J}(x_1,x_2)$ and
$\tens{J'}(x'_1,x'_2)$, where $\tenscp{J}_{ij} = \partial y_i/\partial x_j$ 
and $\tenscp{J}'_{ij} = \partial y'_i/\partial x'_j$, are related by 
\begin{equation}
\tens{J'} = {\bmath{\mathcal R}}^\rmn{T}\,\tens{J}\,{\bmath{\mathcal R}}\,,
\label{eq:hat}
\end{equation}
so that $\tens{J'}$ becomes diagonal. 
In analogy to Eq.~(\ref{eq:phiabbr}), let indices to $\Phi'$ denote 
derivatives with respect to $x'_i$.
The invariance of trace and determinant under the transformation given by
Eq.~(\ref{eq:hat}) yields
\begin{equation}
\Phi'_{11} = \lambda^{(1)} = \Phi_{11} + \Phi_{22} \,,\quad
\Phi'_{22} = \lambda^{(2)} = 0\,.
\label{eq:phi11}
\end{equation}

In the chosen coordinates, the gradient of the Jacobian determinant reads
\begin{equation}
{\vec \nabla}_{\vec x} \det \tens{J}' = \Phi'_{11}\,\left(\begin{array}{c}
\Phi'_{122} \\ \Phi'_{222} \end{array}\right)\,,
\label{eq:graddetJp}
\end{equation}
so that a tangent vector to the critical curve has to be proportional to
$(-\Phi'_{222},\Phi'_{122})^{\rmn{T}}$.
Therefore, the 
conditions for a fold become
\citep{SchneiWei:cusp}
\begin{itemize}
\item $\Phi'_{11} \neq 0$, $\Phi'_{22} = 0$,
\item $\Phi'_{222} \neq 0$,
\end{itemize}
which guarantee that the gradient of the Jacobian determinant does not vanish.
Taking into account these conditions and the specific choice of coordinates, the 
lens equation near a fold can be expanded as
\citep[][p.~183f.]{SEF}
\begin{eqnarray}
y_1' & = & \Phi'_{11} x'_1 + \frac{1}{2}\,\Phi'_{122} (x'_2)^2 + \Phi'_{112} x'_1 x'_2 
\label{eq:y1pf}\,,  \\
y_2' & = & \frac{1}{2}\,\Phi'_{112} (x'_1)^2 + \Phi'_{122} x'_1 x'_2 +
\frac{1}{2}\,\Phi'_{222} (x'_2)^2\,. \label{eq:y2pf}
\end{eqnarray}
Using this expansion, the Jacobian matrix near the fold reads
\begin{equation}
\tens{J}' = \left(\begin{array}{cc}
\Phi'_{11}+\Phi'_{112} x'_2 & \Phi'_{112} x'_1 + \Phi'_{122} x'_2 \\
\Phi'_{112} x'_1 + \Phi'_{122} x'_2 & \Phi'_{122} x'_1 + \Phi'_{222} x'_2
\end{array}\right)\,,
\end{equation}
and to lowest order its determinant 
becomes
\begin{equation}
\det \tens{J}' = \Phi'_{11} (\Phi'_{112} x'_1+ \Phi'_{222} x'_2)\,,
\label{eq:detJfold}
\end{equation}
in accordance with Eq.~(\ref{eq:graddetJp}).
The condition $\det \tens{J}'({{\vec x}'}^{(\rmn{crit})}) = 0$
yields the
critical curve as 
$\Phi'_{112}\, {x'}^\rmn{(crit)}_1+ \Phi'_{222}\, {x'}^\rmn{(crit)}_2 = 0$,
so that with Eqs.~(\ref{eq:y1pf}) and~(\ref{eq:y2pf}),  
the caustic is obtained (to lowest order) as the parabola
\begin{equation}
{y'}^\rmn{(crit)}_2 = \frac{\Phi'_{112}\Phi'_{222}-(\Phi'_{122})^2}{2 (\Phi'_{11})^2\, \Phi'_{222}}\,
({y'}^\rmn{(crit)}_1)^2\,.
\end{equation}
For source positions along its symmetry axis ($y'_1 = 0$), 
Eqs.~(\ref{eq:y1pf}) and~(\ref{eq:y2pf})
yield two 
images 
\begin{eqnarray}
x'_1 & = & -\frac{\Phi'_{122}}{\Phi'_{11} \Phi'_{222}}\,y'_2\,, \label{eq:xp1img}  \\
x'_2 & = & \pm \,\sqrt{\frac{2 y'_2}{\Phi'_{222}}} \label{eq:xp2img}
\end{eqnarray}
for $y'_2 > 0$ which degenerate into a single (critical image) for $y'_2 = 0$, while there are
no images for $y'_2 < 0$. It follows that the
normal to the caustic pointing to the inside is in the positive $y'_2$-direction, i.e.\
${\vec n}'_\rmn{f} = (0,1)^{\rmn{T}}$, so that in $(y_1,y_2)$-coordinates,
${\vec n}_\rmn{f} = {\vec e}^{(2)}$ is the 
inside-pointing caustic normal at ${\vec y}_\rmn{f}$.
Neglecting $y'_2$ against $\sqrt{y'_2}$, Eqs.~(\ref{eq:detJfold}), 
(\ref{eq:xp1img}), and~(\ref{eq:xp2img})
yield the magnification due to the critical images as 
\begin{equation}
A_\rmn{crit}(0,y'_2) = \sqrt{\frac{R_\rmn{f}}{y'_2}}\,\Theta(y'_2)\,,
\label{eq:Acritgen}
\end{equation}
where
\begin{equation}
R_\rmn{f} = \frac{2}{(\Phi'_{11})^2\,|\Phi'_{222}|} > 0
\label{eq:calcrf}
\end{equation}
measures the strength of the strength of the fold caustic and
can also be interpreted as a characteristic distance scale.

The relation given by Eq.~(\ref{eq:rder}) implies that
derivatives of the Fermat-potential $\Phi$ with respect to $x'_i$ and $x_i$ are
related by 
\begin{eqnarray}
\Phi'_{ij} & = &  \sum_{k,l = 1}^2 \Phi_{kl}\,
{{\mathcal R}}_{ki}\,{{\mathcal R}}_{lj}\,,\\
\Phi'_{ijk} & = & \sum_{l,m,n = 1}^2 \Phi_{lmn}\,{{\mathcal R}}_{li}
\,{{\mathcal R}}_{mj}\,{{\mathcal R}}_{nk}\,, 
\end{eqnarray}
the first of the equations being equivalent with Eq.~(\ref{eq:hat}).
Explicitly, the most relevant derivatives for a fold caustic read
\begin{eqnarray}
\Phi'_{11} & = &  \cos^2\theta\,\Phi_{11}+ 2\,\sin\theta\,\cos\theta\,
\Phi_{12} + \sin^2\theta\,\Phi_{22}\,,\\
\Phi'_{222} & = & \cos^3\theta\,\Phi_{111} 
+3\,\sin^2\theta\,\cos\theta\,
\Phi_{112} \;- \nonumber \\ & & \quad -\;3\,\sin\theta\,\cos^2\theta\,\Phi_{122}+ 
\cos^3 \theta\,\Phi_{222}\,.
\end{eqnarray}
By inserting the explicit values for $\sin \theta$ and $\cos \theta$ given by
Eq.~(\ref{eq:sincosexpl}), one reveals 
$\Phi'_{11} = \Phi_{11} + \Phi_{22}$, Eq.~(\ref{eq:phi11}).

\section{Magnification of source stars}
\label{sec:magstar}

The magnification of a point source in the vicinity of a fold caustic can be decomposed 
into the contributions by the two critical images and the remaining images, i.e.\
\begin{equation}
A^\rmn{p}_\rmn{fold}({\vec y}) = A^\rmn{p}_\rmn{crit}({\vec y}) + 
A^\rmn{p}_\rmn{other}({\vec y})\,.
\end{equation}
If one neglects the curvature of the caustic and any changes of the lens properties in the
direction perpendicular to it, Eq.~(\ref{eq:Acritgen}) together with the relation
$\det \tens{J} = \det \tens{J}'$ yields the magnification for a point source
at $\vec y$ due to the critical images as
\begin{equation}
A^\rmn{p}_\rmn{crit}(y_\perp; R_\rmn{f}) = 
\biggl(\frac{y_\perp}{R_\rmn{f}}\biggr)^{-1/2}\,\Theta(y_\perp)
\,,
\label{eq:afoldpt}
\end{equation}
where 
$y_\perp  = \left({\vec y} - {\vec y}_\rmn{f}\right)\cdot {\vec n}_\rmn{f}$
is the distance perpendicular to the caustic 
(c.f.\ \citealt{SchneiWei:twomass}; \citealt[][p.186f.]{SEF};
\citealt{PLANET:sol}). 
For the validity and limits of this approximation, the reader
is referred to \citet{GaudiPetters:fold}.

Let us consider a circular source star with radius $\rho_\star$ and surface
brightness
$I^{(s)}(\rho) = {\overline I}^{(s)}
\,\xi^{(s)}(\rho)$ 
for a filter $s$ 
as function of the fractional radius $0 \leq \rho \leq 1$,
where ${\overline I}^{(s)}$ denotes the average value,
and $\xi^{(s)}(\rho)$ is a dimensionless function describing the radial 
stellar brightness profile,
which is normalized to yield unity if integrated over the source, i.e.\
\begin{equation}
\int\limits_0^1 \xi^{(s)}(\rho)\,\rho\,\mbox{d}\rho = \frac{1}{2}\,.
\label{eq:ldnorm}
\end{equation}
With $I^{(s)}(\rho)$ being composed of the specific intensities 
$I^\lambda(\rho)$ for the wavelength $\lambda$ 
by means of the transmission function
$B^{(s)}(\lambda)$ as
\begin{equation}
I^{(s)}(\rho) = \int\limits_0^\infty I^\lambda(\rho)\,B^{(s)}(\lambda)\,d\lambda\,,
\end{equation}
and $I^\lambda(\rho) = \overline{I}^\lambda\,\xi^\lambda(\rho)$, where
$\overline{I}^\lambda$ denotes the average intensity and
$\xi^\lambda(\rho)$ a dimensionless normalized 
profile function for the specific wavelength,
one obtains
\begin{equation}
\xi^{(s)}(\rho) = \frac{\int\limits_0^\infty
 \overline{I}^\lambda B^{(s)}(\lambda)\,
 \xi^\lambda(\rho)\, \rmn{d}\lambda}
{\int\limits_0^\infty
 \overline{I}^\lambda B^{(s)}(\lambda)\,
 \rmn{d}\lambda}
\,.
\end{equation}

For the source 
being centered at ${\vec y}$, one obtains
the magnification by convolving 
Eq.~(\ref{eq:afoldpt})
with the brightness profile $\xi^{(s)}(\rho)$, yielding
\begin{eqnarray}
& & \hspace*{-2em} A_\rmn{crit}(y_\perp, \rho_\star; R_\rmn{f};\xi^{(s)}) = 
A_\rmn{crit}^{0}(y_\perp/\rho_\star, \rho_\star/R_\rmn{f};
\xi^{(s)}) \nonumber \\
& & =
\biggl(\frac{R_\rmn{f}}{\rho_\star}\biggr)^{1/2}\, 
G^\star_\rmn{f}\biggl(1+\frac{y_\perp}{\rho_\star};\xi^{(s)}\biggr)\,,
\label{eq:foldpat}
\end{eqnarray}
where $G^\star_\rmn{f}$ is a universal fold-caustic profile function depending
on the brightness profile $\xi^{(s)}$ of the source given by 
\citep{SchneiWei:QSO}
\begin{eqnarray}
& & \hspace*{-2em} G^\star_\rmn{f}(\eta; \xi^{(s)}) =  
\frac{2}{\upi}\,\int\limits_{\max(1-\eta,-1)}^{\max(1-\eta,1)} \frac{1}{\sqrt{\rho_\perp+\eta-1}}\,\times \nonumber\\
 & & \times\,
\int\limits_0^{\sqrt{1-\rho_\perp^2}}
\xi^{(s)}\left(\sqrt{\rho_\perp^2+\rho_\parallel^2}\right)\,
\mbox{d}\rho_\parallel\,\mbox{d}\rho_\perp\,.
\label{eq:univldprofile}
\end{eqnarray}
In contrast to several previous papers, $\eta$ is defined so that 
the source is completely outside the caustic for $\eta < 0$,
where $G^\star_\rmn{f}(\eta;\xi^{(s)}) = 0$, completely inside the caustic for
$\eta > 2$, and its center crosses the caustic for
$\eta = 1$. 
As can be seen directly from Eq.~(\ref{eq:foldpat}),
$A_\rmn{crit}(y_\perp, \rho_\star; R_\rmn{f}; \xi^{(s)})$ depends
only on the two ratios $y_\perp/\rho_\star$ and $\rho_\star/R_\rmn{f}$.
Explicit forms of $G^\star_\rmn{f}$ for selected
brightness profiles $\xi^{(s)}$ will be discussed in Sect.~\ref{sec:magprofiles}.

Compared to $A^\rmn{p}_\rmn{crit}$, 
the magnification due to the remaining, non-critical images is
varying slowly in the vicinity of a caustic, so that it can be expanded as
\begin{equation}
A^\rmn{p}_\rmn{other}({\vec y}) = A_\rmn{f} + ({\vec y}-{\vec y}_\rmn{f})\cdot
({\vec \nabla} A)_\rmn{f}\,,
\label{eq:Aotherexpand}
\end{equation}
where $A_\rmn{f} = A^\rmn{p}_\rmn{other}({\vec y}_\rmn{f})$ and
$({\vec \nabla} A)_\rmn{f} = {\vec \nabla} A^\rmn{p}_\rmn{other}({\vec y}_\rmn{f})$.
It is assumed that $({\vec \nabla} A)_\rmn{f} \neq 0$, while otherwise the use of 
higher-order terms in the expansion would be indicated.\footnote{The condition 
$({\vec \nabla} A)_\rmn{f} \neq 0$ is thought to hold for all fold singularities of 
binary lenses, since
$({\vec \nabla} A)_\rmn{f} = 0$ would require a symmetry of $A_\rmn{f}$
to spatial variations in directions both parallel and perpendicular to the line connecting
the two lens objects, which should only be achievable for source positions in between the lens 
objects, where however a saddlepoint rather than a maximum or minimum is expected.}
Both $A_\rmn{f}$ and $({\vec \nabla} A)_\rmn{f}$ can be expressed by means of derivatives
of the Fermat potential $\Phi$ evaluated at the $\tilde N$ non-critical images ${\vec x}^{(i)}$ 
of the source at ${\vec y}_\rmn{f}$. As outlined in detail by \citet{WM95:fifth}, the lens
equation for a binary lens can be written as 5th-order complex polynomial in 
$z = x_1 + \rmn{i}\,x_2$
which can be solved numerically
for the  
the images $z^{(i)} = x^{(i)}_1 + \rmn{i}\,x^{(i)}_2$ by standard root-finding routines.  
While $A_\rmn{f}$ is simply given by
\begin{equation}
A_\rmn{f} = \sum_{i=1}^{\tilde N} 
\frac{1}{\left|\Phi_{11} \Phi_{22} - (\Phi_{12})^2\right|}\,,
\label{eq:calcAf}
\end{equation}
the gradient $({\vec \nabla} A)_\rmn{f}$ can be determined as follows. 

For an image at ${\vec x}_0$ and the corresponding source at ${\vec y}_0$, derivatives
with respect to the components of ${\vec x}$ and ${\vec y}$ are related by 
\begin{equation}
{\vec \nabla}_{\vec y} = \tens{J}^{-1}({\vec x}_0)\,{\vec \nabla}_{\vec x}\,,
\end{equation}
so that 
$({\vec \nabla} A)_\rmn{f}$
reads 
\begin{eqnarray}
 & & \hspace*{-2em}({\vec \nabla} A)_\rmn{f}  =  \sum_{i=1}^{\tilde N}
\tens{J}^{-1}({\vec x}^{(i)})\,{\vec \nabla}_{\vec x} |\mu({\vec x}^{(i)})| \nonumber \\
 & & = - \sum_{i=1}^{\tilde N} \frac{\mbox{sign}(\det \tens{J}({\vec x}^{(i)}))}
{\left|\det \tens{J}({\vec x}^{(i)})\right|^2}\,
\tens{J}^{-1}({\vec x}^{(i)})\,{\vec \nabla}_{\vec x} \det \tens{J}({\vec x}^{(i)})\,.
\end{eqnarray}
With
\begin{equation}
\tens{J}^{-1} = \frac{1}{\det \tens{J}}\,\left(\begin{array}{cc}
\Phi_{22} & -\Phi_{12} \\ - \Phi_{12} & \Phi_{11} \end{array}\right)
\end{equation}
and
\begin{equation}
{\vec \nabla}_{\vec x} \det \tens{J} = \left(\begin{array}{c}
\Phi_{11} \Phi_{122} - 2\,\Phi_{12} \Phi_{112} + \Phi_{22} \Phi_{111} \\  
\Phi_{11} \Phi_{222} - 2\,\Phi_{12} \Phi_{122} + \Phi_{22} \Phi_{112} \end{array}\right)\,,  
\end{equation}
$({\vec \nabla} A)_\rmn{f}$ expressed by means of the derivatives of the Fermat
potential $\Phi$ becomes
\begin{eqnarray}
 & & \hspace*{-2em}({\vec \nabla} A)_\rmn{f}  =
- \sum_{i=1}^{\tilde N} \frac{1}{\left|\Phi_{11} \Phi_{22} - (\Phi_{12})^2\right|^3}\,\times \nonumber \\
& & 
 \times\,
\left(\begin{array}{l}
-\Phi_{11} \Phi_{22} \Phi_{222}\,+\\ \qquad\qquad
+\,[\Phi_{11}\Phi_{22}+2\,(\Phi_{12})^2]\,\Phi_{122}\,- \\
\qquad \qquad -\, 3\,\Phi_{12}\Phi_{22}\Phi_{112}+\Phi_{22}^2\Phi_{111} \\
\Phi_{11}^2\Phi_{222} - 3\,\Phi_{11}\,\Phi_{12}\,\Phi_{122}\, + \\
\qquad \qquad +\, [\Phi_{11}\Phi_{22}+ 2\,(\Phi_{12})^2]\,\Phi_{112}\,-\\ 
\qquad \qquad -\, \Phi_{12}\Phi_{22}\Phi_{111}
\end{array}\right)\,.
\label{eq:calcnablaAf}
\end{eqnarray}

Due to symmetry, the magnification of 
an extended circular source $A_\rmn{other}({\vec y},\rho_\star;
\xi^{(s)})$, obtained by integration of the
convolution of the magnification $A^\rmn{p}_\rmn{other}$ 
as given by Eq.~(\ref{eq:Aotherexpand}) with the radial source brightness profile, turns out to be  
equal to that of a point source at its center, i.e.\
$A_\rmn{other}({\vec y},\rho_\star;
\xi^{(s)}) = A^\rmn{p}_\rmn{other}({\vec y})$
unless the source size exceeds
the range for which the expansion is a fair approximation, so that the total magnification
of an extended circular source reads 
$A_\rmn{fold}({\vec y},\rho_\star;\xi^{(s)}) =
A_\rmn{crit}({\vec y},\rho_\star;\xi^{(s)}) 
+ A^\rmn{p}_\rmn{other}({\vec y})$.

\section{Limb darkening}

\label{sec:magprofiles}

With $\vartheta$ denoting
the angle between the normal to the stellar surface and the direction to the observer,
so that $\cos \vartheta = \sqrt{1-\rho^2}$,
normalized brightness profile functions
\begin{equation}
\xi^{(s)}_{\{p\}}(\rho) = (1+p/2)\,(1-\rho^2)^{p/2}
\end{equation}
are proportional to $\cos^p \vartheta$, and their linear
superposition
\begin{equation}
\xi^{(s)}(\rho; \Gamma^{(s)}_{\{p_1\}}\ldots \Gamma^{(s)}_{\{p_{k_\rmn{I}}\}}) = \
1+\sum_{i=1}^{k_\rmn{l}} \Gamma^{(s)}_{\{p_i\}}\,\left[\xi^{(s)}_{\{p_i\}}(\rho)-1\right]\,,
\end{equation}
where $p_i > 0$, provides a popular class of models for limb darkening.
The contribution of individual base profiles $\xi^{(s)}_{\{p_i\}}(\rho)$ to the 
full brightness profile function $\xi^{(s)}(\rho)$ is measured by the $k_\rmn{l}$ limb-darkening
coefficients $0 \leq \Gamma^{(s)}_{\{p_i\}} \leq 1$, which moreover have to fulfill the
condition
\begin{equation}
\sum_{i=1}^{k_\rmn{l}} \Gamma^{(s)}_{\{p_i\}} \leq 1\,.
\end{equation}

\begin{figure}
\includegraphics[width=84mm]{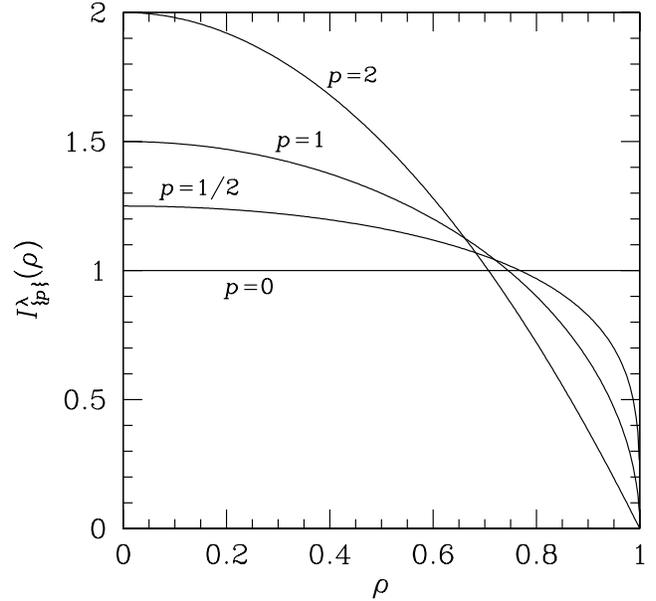}
\caption{Stellar brightness profiles $\xi^{(s)}_{\{p\}}(\rho) = (1+p/2)\,(1-\rho^2)^{p/2}$ as
a function of the fractional radius $\rho$ for source stars with 
uniform brightness ($p=0$),
square-root ($p=1/2$), linear ($p=1$), and quadratic ($p=2$) limb darkening.}
\label{fig:ldlaws}
\end{figure}

The stellar brightness profiles $\xi^{(s)}_{\{p\}}$ 
for uniform brightness ($p=0$), square-root ($p=1/2$), linear ($p=1$) and
quadratic limb darkening ($p=2$)
as a function of the fractional radius
$\rho$ are shown in Fig.~\ref{fig:ldlaws}. In general, larger powers of $p$ provide
stronger limb darkening.

By inserting the brightness profile $\xi^{(s)}_{\{p\}}(\rho)$ into Eq.~(\ref{eq:univldprofile}), one obtains
its fold-caustic magnification 
function $G^{\star}_{\rmn{f},\{p\}}(\eta) \equiv G^{\star}_{\rmn{f}}(\eta; \xi^{(s)}_{\{p\}})$ as
\citep{SchneiWei:QSO}
\begin{equation} 
G^{\star}_{\rmn{f},\{p\}}(\eta) = \frac{1}{\sqrt{\upi}}\,
\frac{(1+\frac{p}{2})!}{(\frac{1+p}{2})!}\,
\int\limits_{\rm max(1-\eta,-1)}^{\rm max(1-\eta,1)}\frac{(1-x^2)^{\frac{1+p}{2}}}{\sqrt{x+\eta-1}}\;
\rmn{d}x\,.
\label{eq:Gfp}
\end{equation}

\begin{figure}
\includegraphics[width=84mm]{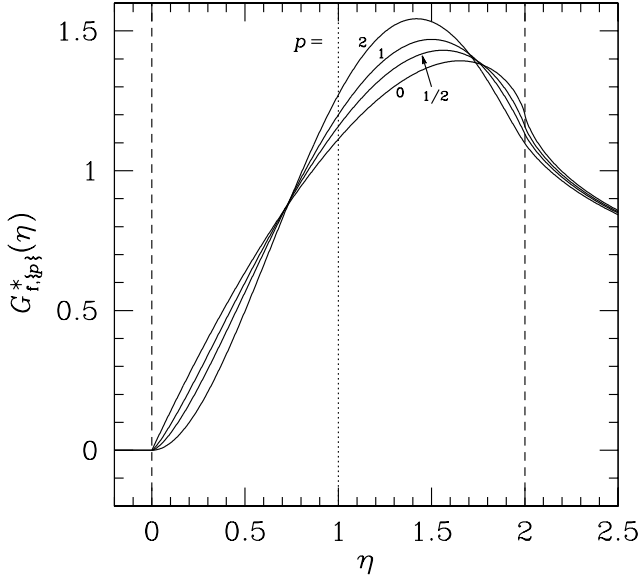}
\caption{Universal fold-caustic magnification function 
$G^\star_{\rmn{f},\{p\}}(\eta)$ for
selected source brightness profiles following power-laws 
in $\cos \vartheta = \sqrt{1-\rho^2}$ of the form
$\xi^{(s)}_\rmn{\{p\}}(\rho) = (1+p/2)\,(1-\rho^2)^{p/2}$: 
Uniform brightness ($p=0$),
square-root ($p=1/2$), linear ($p=1$), and quadratic ($p=2$) limb darkening.}
\label{fig:magld}
\end{figure}

For even $p$ (including $p=0$), the fold-caustic magnification 
function $G^{\star}_{\rmn{f},\{p\}}$ can be expressed
by means of the complete elliptical integrals of the first and second kind, $K(x)$ and $E(x)$, respectively
\citep[e.g.\ ][]{GradRyzh}, while $G^{\star}_{\rmn{f},\{p\}}$ becomes an analytical function for odd $p$.  
For uniformly bright sources ($p=0$), one obtains \citep{SchneiWei:QSO}
\begin{equation}
G^{\star}_{\rmn{f},\{0\}}(\eta) = \left\{\begin{array}{l}
0 \hfill \mbox{for} \quad \eta \leq 0 \\
\hspace*{-0.1em}
\frac{4\,\sqrt{2}}{3 \upi}\left[(2-\eta)\,
K\left(\sqrt{\frac{\eta}{2}}\,\right)\right.- \\  \quad\left. - 2\,(1-\eta)\,
E\left(\sqrt{\frac{\eta}{2}}\,\right)\right] \\  
\qquad \qquad \qquad \qquad \qquad \hfill \mbox{for} \quad 
 0 < \eta < 2 \\
\hspace*{-0.1em}
\frac{8}{3 \upi}\sqrt{\eta}\left[(2-\eta)\,
K\left(\sqrt{\frac{2}{\eta}}\,\right)\right.- \\ \quad  \left.  -  (1-\eta)\,
E\left(\sqrt{\frac{2}{\eta}}\,\right)\right]  \hfill \mbox{for} \quad 
 \eta \geq 2 \end{array}\right.\,,
\end{equation}
while for linear limb darkening ($p = 1$) the evaluation of Eq.~(\ref{eq:Gfp}) yields 
\citep{SchneiWag}
\begin{equation}
G^{\star}_{\rmn{f},\{1\}}(\eta) = \left\{\begin{array}{l}
0 \qquad \qquad \hfill \mbox{for} \quad \eta \leq 0 \\
\frac{2}{5}\,(5-2 \eta)\,\eta^{3/2} \qquad \qquad \hfill \mbox{for} \quad
 0 < \eta \leq 2 \\
\frac{2}{5}\,\left[(5-2 \eta)\,\eta^{3/2}\right.\,+ \\
 \quad +\,\left.(1+2 \eta)\,(\eta-2)^{3/2}\right] \;  \hfill \mbox{for} \quad
 \eta > 2 \\
\end{array}\right.\!\!\!\,.
\end{equation}
The fold-caustic magnification 
functions $G^{\star}_{\rmn{f},\{p\}}$ for uniformly bright sources ($p=0$),
square-root ($p=1/2$), linear ($p=1$) and quadratic limb darkening ($p=2$) 
are shown in Fig.~\ref{fig:magld}.
With the stellar surface brightness vanishing at the limb for all
power-law profiles with $p>0$, the slope of $G^{\star}_{\rmn{f},\{p\}}$
becomes zero when the stellar limb touches the caustic from the outside  
($\eta = 0$), In contrast, there is a slope discontinuity at this
phase for $p=0$ and therefore for all brightness profiles that involve a
non-zero constant brightness term. 
As the source enters the caustic, the magnification rises to a
peak and thereafter
falls asymptotically
with the inverse square root of its perpendicular distance to the caustic.
The position of the peak depends on the brightness profile and occurs
for the source center being inside
the caustic.
With more light being concentrated near the source center, stronger limb darkening 
can account for fold-caustic peak magnifications similar to that for less
limb-darkened but smaller sources.
On entering the caustic, stronger limb darkening produces less steep initial rises
of the magnification, while
steeper rises occur before a narrower peak at higher magnification
is reached for a smaller distance of the
source center from the caustic. 
After the peak, the magnification for extended sources
exceeds that of a point source, where the difference increases for weaker limb darkening.  

For more details about the fold-caustic magnification function $G^\star_{\rmn{f},\{p\}}$, the reader
is referred to \citet{Rhie:ld}, while more about the extraction of limb-darkening
parameters from the observed data during a fold-caustic passage event can be found
in two recent papers by \citet{Do:ld,Do:2ndld}.

\section{Lightcurves near folds}
\label{sec:parametrization}

Let the center of the source cross the caustic at time $t_\rmn{f}$ and let
$t^\star_\rmn{f}$ denote the point of time when the source begins to enter
or finishes to exit the caustic. If the source brightness does not 
vanish at the stellar limb,
$G^\star_\rmn{f}(\eta;\xi^{(s)})$
involves a slope discontinuity at $\eta = 0$, producing a characteristic feature
in the lightcurve at time $t^\star_\rmn{f}$, whereas $t_\rmn{f}$ is much
less easily recognizable from the lightcurve. The choice of $t^\star_\rmn{f}$
as reference point rather than $t_\rmn{f}$ constitutes an importance 
difference to the discussion presented by \citet{PLANET:sol} and is the
key point for avoiding
strong correlations between model parameters.
With $t_\star^\perp = \pm\,(t_\rmn{f} - t^\star_\rmn{f}) \geq 0$ (throughout the paper,
the upper sign will refer to a caustic entry,
while the lower sign will refer to a caustic exit),
the source crosses the caustic during the timespan $2\,t_\star^\perp$.
Let $0 < \phi < \upi$ 
denote the angle from the caustic tangent to the source trajectory,
and $t_\rmn{E} > 0$
denote the timespan in which the source moves by a unit distance.
During the time $t_\rmn{E}^\perp = t_\rmn{E}/(\sin \phi)$, 
the source therefore moves by a unit distance perpendicular to the caustic,
and $t_\star^\perp = \rho_\star\,t_\rmn{E}^\perp$, so that
\begin{equation}
y_\perp = 
\pm\, \frac{t-t_\rmn{f}}{t_\rmn{E}^\perp} = 
\pm\, \rho_\star\,\frac{t-t_\rmn{f}}{t_\star^\perp} = 
\rho_\star\,\left(\pm\,\frac{t-t^\star_\rmn{f}}{t_\star^\perp}-1\right)\,.
\end{equation}
With ${\hat t}$ denoting an arbitrarily chosen unit time, and 
\begin{equation}
t_\rmn{r} \equiv R_\rmn{f}\,t_\rmn{E}^\perp > 0
\end{equation}
a characteristic caustic rise time, one can define a caustic rise parameter
\begin{equation}
\zeta_\rmn{f} \equiv \sqrt{\frac{t_\rmn{r}}{\hat t}} > 0\,.
\end{equation}
With these definitions, Eq.~(\ref{eq:foldpat}) yields the magnification due to the critical images
as
\begin{eqnarray}
A_\rmn{crit}(t) & \equiv & 
A_\rmn{crit}^{0}\left(\pm\,\frac{t-t^\star_\rmn{f}}{t_\star^\perp}-1,
\frac{t_\star^\perp}{\zeta_\rmn{f}^2\,\hat t};\xi^{(s)}\right) 
\nonumber \\
& = & {\zeta_\rmn{f}}\,\left(\frac{\hat t}{t_\star^\perp}\right)^{1/2}\, 
G^\star_\rmn{f}\biggl(\pm\,\frac{t-t^\star_\rmn{f}}{t_\star^\perp};\xi^{(s)}\biggr)\,.
\end{eqnarray}

The singularity in this expression for $t_\star^\perp = 0$ (point source) can be avoided
by defining a function
\begin{equation}
{\hat G}^\star_\rmn{f}\left({\hat y}_\perp^\star,
{\hat \rho}_\star^\perp;
\xi^{(s)}\right)  = 
\left\{\begin{array}{l}
\left({\hat \rho}_\star^\perp\right)^{-1/2}\,\times \\ \quad \times\,
G^\star_\rmn{f}\left({\hat y}_\perp^\star/{\hat \rho}_\star^\perp;
\xi^{(s)}\right) \\ \qquad \qquad \hfill \mbox{for} \quad 
{\hat \rho}_\star^\perp \neq 0 \\
\left({\hat y}_\perp^\star\right)^{-1/2}\,
\Theta\left({\hat y}_\perp^\star\right) \\
\hfill \mbox{for} \quad {\hat \rho}_\star^\perp = 0\end{array}\,,
\right.
\end{equation}
which is continuous
in $t_\star^\perp = 0$, i.e.~
\begin{eqnarray}
& & \hspace*{-2em}
\lim_{t_\star^\perp \to 0} {\hat G}^\star_\rmn{f}\left(\pm\,\frac{t-t^\star_\rmn{f}}{\hat t},
\frac{t_\star^\perp}{\hat t};\xi^{s)}\right)\, = \nonumber \\ 
 & & =\,
{\hat G}^\star_\rmn{f}\left(\pm\,\frac{t-t^\star_\rmn{f}}{\hat t},0;\xi^{(s)}\right)\,,
\end{eqnarray}
and allows the magnification due to the critical images
to be expressed as
\begin{equation}
A_\rmn{crit}(t) = 
 \zeta_\rmn{f}\, 
{\hat G}^\star_\rmn{f}\biggl(\pm\,\frac{t-t^\star_\rmn{f}}{\hat t},
\frac{t_\star^\perp}{\hat t}; \xi^{(s)}\biggr)\,.
\end{equation}

Consider $n$ lightcurves being observed and let
$F_\rmn{S}^{(s)} > 0$ and $F_\rmn{B}^{(s)}$ denote the fluxes of the source 
and the background, respectively, for the $s$-th lightcurve,
so that the
blend ratio is given by $g^{(s)} = F_\rmn{B}^{(s)}/F_\rmn{S}^{(s)}$ and
the baseline flux is $F_\rmn{base}^{(s)} = 
F_\rmn{S}^{(s)} + F_\rmn{B}^{(s)}$.
In analogy to the spatial expansion of the magnification 
$A^\rmn{p}_\rmn{other}$ due to the non-critical images 
as given by Eq.~(\ref{eq:Aotherexpand}), 
its temporal expansion around
$t^\star_\rmn{f}$ reads
\begin{equation}
A^\rmn{p}_\rmn{other}(t)
 \simeq 
A^\star_\rmn{f}+  (t-t^\star_\rmn{f}) \dot{A}^\star_\rmn{f}\,,
\label{eq:expandafoldoth}
\end{equation}
where $A^\star_\rmn{f} = 
A^\rmn{p}_\rmn{other}(t^\star_\rmn{f})$ and
$\dot{A}^\star_\rmn{f} = 
\dot{A}^\rmn{p}_\rmn{other}(t^\star_\rmn{f})$.
The relations between the 
coefficients $A^\star_\rmn{f}$, $\dot{A}^\star_\rmn{f}$,
$A_\rmn{f}$,    
and $({\vec \nabla} A)_\rmn{f}$ will be
discussed In Section~\ref{sec:fullmodel}.

Let us define
\begin{equation}
F_\rmn{r}^{(s)} 
\equiv \zeta_\rmn{f}\,F_\rmn{S}^{(s)} > 0\,,
\label{eq:defFr}
\end{equation}
\begin{equation}
{\hat \omega}^\star_\rmn{f} \equiv \pm\,
\frac{\dot{A}^\star_\rmn{f}}{\zeta_\rmn{f}}\,,
\label{eq:defomega}
\end{equation}
and the flux at time $t^\star_\rmn{f}$ as
\begin{equation}
F^{\star(s)}_\rmn{f} \equiv  F_\rmn{S}^{(s)}\,A^\star_\rmn{f} +
F_\rmn{B}^{(s)} = F_\rmn{S}^{(s)}\,(A^\star_\rmn{f}+g^{(s)})\,.
\label{eq:fluxfold}
\end{equation}

Using the above definitions,
the 
total flux $F^{(s)}_\rmn{fold}(t) = F_\rmn{S}^{(s)}\,
\left[A_\rmn{crit}(t)+
A^\rmn{p}_\rmn{other}(t)\right]
+ F_\rmn{B}^{(s)}$ for the $s$-th lightcurve takes the form 
\begin{eqnarray}
& & \hspace*{-2em} F^{(s)}_\rmn{fold}(t) = F_\rmn{r}^{(s)}\,\left[
{\hat G}^\star_\rmn{f}\left(\pm\,\frac{t-t^\star_\rmn{f}}{\hat t},
\frac{t_\star^\perp}{\hat t};\xi^{(s)}\right)\right.\,\pm \nonumber \\
& & \qquad \qquad \pm\,
{\hat \omega}^\star_\rmn{f}\,(t-t^\star_\rmn{f})\bigg] 
+ F^{\star(s)}_\rmn{f} 
\,.
\label{eq:ffgen}
\end{eqnarray}
A fit to the $n$ 
observed lightcurves therefore involves the $3+2n$ parameters
$t^\star_\rmn{f}$, $t_\star^\perp > 0$,  $F_\rmn{r}^{(s)} >0$, $F^{\star\,(s)}_\rmn{f}$,
and ${\hat \omega}^\star_\rmn{f}$.
The fluxes $F_\rmn{r}^{(s)}$ measure the asymptotic behaviour of the lightcurve for
source positions
towards the inside of the caustic (i.e.\ for $t \to \pm \infty$) and can be interpreted 
as the flux of the two critical images that are produced at the time $t = t_\rmn{f}
\pm {\hat t} = t^\star_\rmn{f}\pm (t_\star^\perp + {\hat t}\,)$ 
if the extended source is replaced
by a point source at its center, where the chosen point of time corresponds to a unit time
after the source center enters or before it exits the caustic which itself takes place at
$t = t_\rmn{f} \pm t_\star^\perp$.
The parameter ${\hat \omega}^\star_\rmn{f}$ describes the temporal variation of 
$A^\rmn{p}_\rmn{other}$, where $\pm\,{\hat \omega}^\star_\rmn{f}$
measures the rate of the corresponding change of flux in units of $F_\rmn{r}^{(s)}$
with positive time.
Since the change of caustic properties in the direction
parallel to it has been neglected, 
the lightcurve is affected by the transverse motion only.
The observable width of the caustic passage given
by $t_\star^\perp$ is the product of the source size parameter $\rho_\star$ and the
time-scale of transverse motion $t_\rmn{E}^\perp$, whereas neither of the two latter quantities
are observables themselves.

\begin{figure}
\includegraphics[width=84mm]{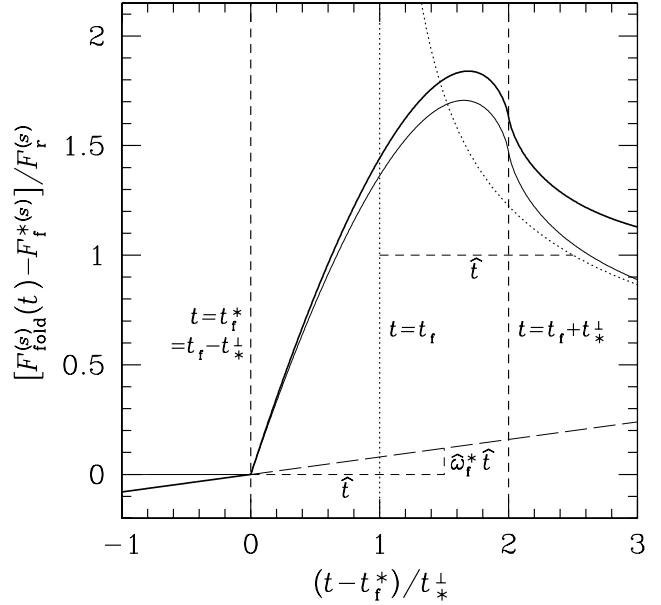}
\caption{Typical lightcurve for a uniformly bright source entering a fold caustic (bold solid line) which
illustrates the physical meaning of the model parameters 
$t^\star_\rmn{f}$, $t_\star^\perp > 0$,  $F_\rmn{r}^{(s)} >0$, $F^{\star(s)}_\rmn{f}$,
and ${\hat \omega}^\star_\rmn{f}$. 
For this example, $t_\star^\perp = 2/3\,{\hat t}$ and ${\hat \omega}_\rmn{f}^\star = 0.08$ have been
chosen.
Also shown are lightcurves with ${\hat \omega}_\rmn{f}^\star = 0$ for the uniformly bright extended source
(light solid line) and the corresponding point source (dotted line). The long-dashed line shows the contribution
by the variation of $A^\rmn{p}_\rmn{other}(t)$.
} 
\label{fig:lcparams}
\end{figure}

The meaning of the model parameters $t^\star_\rmn{f}$, $t_\star^\perp$, 
 $F_\rmn{r}^{(s)}$, $F^{\star\,(s)}_\rmn{f}$,
and ${\hat \omega}^\star_\rmn{f}$ is illustrated in Fig.~\ref{fig:lcparams}.
For observed lightcurves, initial guesses for these parameters can easily be obtained 
from recognizable features.
The point of time $t_\rmn{f}^\star$ is indicated by the arising slope discontinuity, and 
$F^{\star(s)}_\rmn{f}$ is obtained as the flux at $t_\rmn{f}^\star$. The duration of
the caustic passage $2\,t_\star^\perp$ extends approximately between the slope discontinuity
at $t_\rmn{f}^\star$ and the change of sign of the curvature on the other side of the 
caustic peak. With two fluxes $F_1^{(s)} = F_\rmn{fold}^{(s)}(t_1)$ 
and $F_2^{(s)} = F_\rmn{fold}^{(s)}(t_2)$ taken at two points of time $t_1$ and
$t_2$ in the region where the 
source is located inside the caustic and can be fairly
approximated by a point source and the rate of 
change of flux ${\dot F}_\rmn{f}^{\star\,(s)}$ of the non-critical images at 
$t_\rmn{f}^\star$
(obtained as the limit of the tangent for $t \to t_\rmn{f}^\star$ from the outside),
one obtains 
\begin{eqnarray}
F_\rmn{r}^{(s)} & = & \left[(t_1-t_\rmn{f}^\star)\,
\sqrt\frac{\hat t}{\pm(t_2-t_\rmn{f}^\star)-t_\star^\perp}\right.\,- \nonumber \\ 
& & \qquad -\,\left. 
(t_2-t_\rmn{f}^\star)\,
\sqrt\frac{\hat t}{\pm(t_1-t_\rmn{f}^\star)-t_\star^\perp}\right]\nonumber\Bigg/ \\
& & \quad \bigg/ \left[(F_2^{(s)} - F^{\star\,(s)}_\rmn{f})\,(t_1-t_\rmn{f}^\star)\right.\, - \nonumber \\
& & \qquad -\,\left.
(F_1^{(s)} - F^{\star\,(s)}_\rmn{f})\,(t_2-t_\rmn{f}^\star)\right]\,.
\end{eqnarray}
Finally, the slope of any chosen lightcurve $k$ for the source just outside the caustic
denoted by ${\dot F}_\rmn{f}^{\star\,(k)}$ yields 
${\hat \omega}^\star_\rmn{f} = \pm {\dot F}_\rmn{f}^{\star\,(k)}
/F_\rmn{r}^{(k)}$.

In the absence of blending, $F_\rmn{B}^{(s)} = 0$ for all $s$, so that
Eqs.~(\ref{eq:defFr}) and~(\ref{eq:fluxfold}) yield a constant ratio between
$F^{\star(s)}_\rmn{f}$ 
and $F_\rmn{r}^{(s)}$, namely 
\begin{equation}
g_\rmn{f}^\star \equiv \frac{F_\rmn{f}^{\star(s)}}{F_\rmn{r}^{(s)}} = 
\frac{A^\star_\rmn{f}}{\zeta_\rmn{f}} > 0\,.
\label{eq:defgAr}
\end{equation}
In this case, modelling the lightcurve to the observed data
involves $4+n$ independent
parameters, which can be chosen as $t^\star_\rmn{f}$, $t_\star^\perp > 0$, 
$g_\rmn{f}^\star > 0$, $F^{\star(s)}_\rmn{f} > 0$,
and ${\hat \omega}^\star_\rmn{f}$.

While in the case of difference-imaging, 
the flux $F_\rmn{B}^{(s)}$ represents the total background and can
have either sign, it is restricted to $F_\rmn{B}^{(s)} \geq 0$ for
standard photometry. The conditions $F_\rmn{S}^{(s)} > 0$ and
$A_\rmn{f}^\star > 1$ imply $F^{(s)}_\rmn{fold}(t) > 0$ for all $t$ in this 
case, so that
Eq.~(\ref{eq:fluxfold}) requires $F_\rmn{f}^{\star(s)} > 0$.
Therefore, a model
parameter $F^{\star(k)}_\rmn{f} \leq 0$ for the $k$-th lightcurve
implies a negative background flux $F^{(k)}_\rmn{B}$.

For sufficiently small $\pm\,{\hat \omega}^\star_\rmn{f}\,(t-t^\star_\rmn{f})$,
$F^{(s)}_\rmn{fold}$ as given by Eq.~(\ref{eq:ffgen})  
violates the
condition $F^{(s)}_\rmn{fold}(t) > 0$.
Although the approximation of $A_\rmn{other}(t)$ by its expansion around
$t = t^\star_\rmn{f}$ only holds for
 $\left|t-t^\star_\rmn{f}\right| \ll 
\left|A^\star_\rmn{f}/{\dot A}^\star_\rmn{f}\right|$, 
evaluations outside this region may be attempted during a parameter search.
Negative fluxes prohibit the calculation of a corresponding magnitude which
can however be avoided by 
using a modified expansion of $A^\rmn{p}_\rmn{other}$ involving
 an exponential function rather
than a linear function, which is a purely technical modification and has
negligible effect on the lightcurve in the vicinity of the caustic passage.
Let us therefore expand $A^\rmn{p}_\rmn{other}$ as
\begin{equation}
A^\rmn{p}_\rmn{other}(t)
 \simeq 
A^\star_\rmn{f}+  \alpha_\rmn{f}\,
\left(\exp\left\{\frac{\dot{A}^\star_\rmn{f}\,
(t-t^\star_\rmn{f})}{\alpha_\rmn{f}}\right\}-1\right)
\label{eq:expandafoldothexp}\,,
\end{equation}
where $\alpha_\rmn{f} > 0$, so that the local properties
$A^\rmn{p}_\rmn{other}(t^\star_\rmn{f}) = A^\star_\rmn{f}$ and
$\dot{A}^\rmn{p}_\rmn{other}(t^\star_\rmn{f}) 
= \dot{A}^\star_\rmn{f}$ are fulfilled.
 
In analogy to Eq.~(\ref{eq:defgAr}), one can define
\begin{equation}
g_\rmn{f}^{\star\,(s)} \equiv \frac{F_\rmn{f}^{\star(s)}}{F_\rmn{r}^{(s)}} = 
\frac{A^\star_\rmn{f}+g^{(s)}}{\zeta_\rmn{f}} > 0
\end{equation}
and
\begin{equation}
g_{\rmn{f},\rmn{min}}^{\star} \equiv 
\min_{1 \leq s \leq n}\,
\left\{g_\rmn{f}^{\star\,(s)}\right\} > 0\,,
\label{eq:defgmin}
\end{equation}
which corresponds to the lightcurve involving the smallest blend ratio.

With $\alpha_\rmn{f} = \zeta_\rmn{f}\,
g_{\rmn{f},\rmn{min}}^\star$ and 
${\hat \omega}^\star_\rmn{f}$ as defined by Eq.~(\ref{eq:defomega}),
the total flux for the $s$-th lightcurve reads
\begin{eqnarray}
& & \hspace*{-2em}
{\tilde F}^{(s)}_\rmn{fold}(t)  =  F_\rmn{r}^{(s)}\,\left[
{\hat G}^\star_\rmn{f}\left(\pm\,\frac{t-t^\star_\rmn{f}}{\hat t},
\frac{t_\star^\perp}{\hat t};\xi^{(s)}\right)\right.\,+\nonumber \\
& & +\,\left.
g_{\rmn{f},\rmn{min}}^\star\,
\left(\exp\left\{\pm\,\frac{{\hat \omega}^\star_\rmn{f}\,(t-t^\star_\rmn{f})}
{g_{\rmn{f},\rmn{min}}^\star}\right\}-1\right)\right]
+ F^{\star(s)}_\rmn{f} 
\,.
\end{eqnarray}
With the corresponding magnitude being defined as 
\begin{equation}
m_\rmn{fold}^{(s)}(t) = m_\rmn{f}^{\star(s)} -
2.5\,\lg\frac{{\tilde F}^{(s)}_\rmn{fold}(t)}{F_\rmn{f}^{\star(s)}}\,,
\label{eq:defmagfold}
\end{equation} 
where, in analogy to $F_\rmn{f}^{\star(s)}$, 
$m_\rmn{f}^{\star(s)}$ denotes the magnitude at time $t_\rmn{f}^\star$,
the magnitude near a fold-caustic passage takes the form
\begin{eqnarray}
& & \hspace*{-2em}
m_\rmn{fold}^{(s)}(t)  =  m_\rmn{f}^{\star(s)}\, - \nonumber \\
& & -\,2.5\,\lg\left\{
1 + \frac{1}{g_\rmn{f}^{\star\,(s)}}\,
\left[
{\hat G}^\star_\rmn{f}
\left(\pm\,\frac{t-t^\star_\rmn{f}}{\hat t},\frac{t_\star^\perp}{\hat t}
; \xi^{(s)}\right)\right.\right.\,+ \nonumber \\
& & +\,\left.\left.
g_{\rmn{f},\rmn{min}}^\star
\,\left(
\exp\left\{\pm\,\frac{{\hat \omega}^\star_\rmn{f}\,(t-t^\star_\rmn{f})}
{g_{\rmn{f},\rmn{min}}^\star}\right\}-1
\right)\right]
\right\}\,,
\end{eqnarray}
and lightcurve models involve the $3+2n$ parameters
$t^\star_\rmn{f}$, $t_\star^\perp > 0$,
$g_{\rmn{f}}^{\star\,(s)} > 0$, 
$m^{\star(s)}_\rmn{f}$
and ${\hat \omega}^\star_\rmn{f}$
which reduce to the $4+n$ parameters 
$t^\star_\rmn{f}$, $t_\star^\perp > 0$,  
$g_{\rmn{f}}^\star >0$, 
$m^{\star(s)}_\rmn{f}$
and ${\hat \omega}^\star_\rmn{f}$
in the absence of blending.

\section{Full lightcurve and binary lens model}
\label{sec:fullmodel}

\subsection{Constraining the parameter space}
For a galactic microlensing event involving an extended source
at distance $D_\rmn{S}$ and a binary lens with total mass $M$ at distance $D_\rmn{L}$,
let us choose the angular Einstein radius $\theta_\rmn{E}$, defined by Eq.~(\ref{eq:angEinstein}), 
as the unitlength of our coordinates $(x_1,x_2)$ and $(y_1,y_2)$. 
Usually, the resulting lightcurve is characterized by the $7+2n$ model parameters
($d$, $q$, $u_0$, $\alpha$, $t_0$, $t_{\rm E}$, $\rho_{\star}$, $F_{\rm S}^{(s)}$,
$F_{\rm B}^{(s)}$) being described in the following.
The binary lens itself is characterized by 
the angular separation of its components $\delta = 
d\,\theta_\rmn{E}$ and their mass ratio $q$.
The location of the trajectory of the source relative to the lens is described by the closest 
angular separation $\theta_0 = u_0\,\theta_\rmn{E}$ between source and
center of mass of the lens system and the orientation angle $\alpha$ relative to the
line connecting the two lens components. 
The closest approach takes place at time $t_0$ and within the timespan
$t_\rmn{E} = \theta_\rmn{E}/\mu$ the source moves
 by $\theta_\rmn{E}$ relative
to the lens on the sky.
The angular radius of the source is $\theta_\star = \rho_\star\,\theta_\rmn{E}$, and
as defined in Sect.~\ref{sec:parametrization},
$F_{\rm S}^{(s)}$ and $F_{\rm B}^{(s)}$ denote the source and background fluxes,
respectively.

As pointed out by \citet{Do99:CR}, the parameters $d$ and $q$ are usually
strongly correlated resulting in a partial parameter degeneracy which
can be avoided by instead using
the parameter pair 
$(\gamma,q)$ for wide binary lenses, where $\gamma$ denotes 
the shear and the mass ratio $q$ measures the deviation from a Chang-Refsdal lens, or the parameter
pair 
$(\hat Q,q)$ for close binary lenses where $\hat Q$ denotes the absolute value of either
eigenvalue
of the traceless
quadrupole moment and the mass ratio $q$ measures the deviation from a quadrupole lens.
Any $(d,q)$ in the following text may therefore also be understood as
 $(\gamma,q)$ or $(\hat Q,q)$.
 
Rather than by the full set of $7+2n$ model parameters ($d$, $q$, $u_0$, $\alpha$, $t_0$, $t_{\rm E}$, $\rho_{\star}$, $F_{\rm S}^{(s)}$,
$F_{\rm B}^{(s)}$), the lightcurve in the vicinity of a fold caustic 
is completely characterized by $3+2n$ parameters 
$(t_\rmn{f}^\star,t_\star^\perp,F^{(s)}_\rmn{r},F^{\star\,(s)}_\rmn{f},
{\hat \omega}_\rmn{f}^\star)$
or $2+2n$ parameters $(t_\rmn{f}^\star,t_\star^\perp,F^{(s)}_\rmn{r},F^{\star\,(s)}_\rmn{f})$
if one disregards
the parameter ${\hat \omega}_\rmn{f}^\star$ describing the temporal variation of the 
magnification due to non-critical images\footnote{While the inclusion of 
${\hat \omega}_\rmn{f}^\star$ as a model parameter yields more adequate estimates 
for other model parameters,
its value itself is rather uncertain and unreliable.},
as shown in the previous section.
Already a comparison of the number of parameters shows  
that the data over
the caustic passage cannot provide an unique model for the full lightcurve of the event.
Instead, they turn out to be 
insensitive to the caustic crossing angle $\phi$
and the caustic rise parameter $\zeta_\rmn{f}$, as well as to the 
caustic strength $R_\rmn{f}$ and
the magnification $A_\rmn{f}^\star$ at the beginning of the caustic entry or
the end of the caustic exit, the last two parameters given by the 
choice of the singularity ${\vec y}_\rmn{f}$ for a specific binary lens characterized
by $(d,q)$. Therefore, modelling the caustic passage data does not provide measurements
of the time-scale $t_\rmn{E}$, the source and background fluxes  $F_\rmn{S}^{(s)}$ and
$F_\rmn{B}^{(s)}$, the source size parameter $\rho_\star$, and of the characteristics 
of the binary lens, namely the separation parameter $d$ and the mass ratio $q$.

As will be discussed later in this section, there are however some relations 
between and restrictions on these parameters. In particular,
the different rise fluxes $F_\rmn{r}^{(s)}$ provide relations between the 
source fluxes $F_\rmn{S}^{(s)}$, 
and together with the caustic passage fluxes $F^{\star\,(s)}_\rmn{f}$ yield
relations between the background fluxes $F_\rmn{B}^{(s)}$, while 
the parameter ${\hat \omega}_\rmn{f}^\star$ constrains the caustic crossing angle
$\phi$ and relates it to the caustic rise parameter $\zeta_\rmn{f}$.
Moreover, additional constraints apply if measurements of properties of the full lightcurve
such as the time-scale $t_\rmn{E}$, source, background or
baseline fluxes $F_\rmn{S}^{(s)}$, $F_\rmn{B}^{(s)}$, or
$F_\rmn{base}^{(s)}$, or blend ratios $g^{(s)}$ are taken into account as discussed
in Sect.~\ref{sec:constraintsother}.

Nevertheless, with a determination of the angular radius of the source $\theta_\star$, 
e.g.\ from spectral typing, the caustic passage data yield the 
proper motion $\mu^\perp$
perpendicular to the caustic with $t_\star^\perp$ as
\begin{equation}
\mu^\perp = \frac{\theta_\star}{t_\star^\perp}\,,
\end{equation}
whereas the measurement of the full proper motion $\mu$ requires the 
determination of the caustic crossing angle $\phi$ 
\begin{equation}
\mu = \frac{\theta_\star}{t_\star} = 
\frac{\theta_\star}{t_\star^\perp\,\sin \phi} \geq \mu^\perp\,,
\end{equation}
where the caustic passage data only yield a lower limit.

As pointed out by \citet{PLANET:sol},
rather than performing a parameter search on
the $7+2n$-dimensional full parameter set ($d$, $q$, $u_0$, $\alpha$, $t_0$, $t_{\rm E}$, $\rho_{\star}$, $F_{\rm S}^{(s)}$,
$F_{\rm B}^{(s)}$), all suitable binary lens models
can efficiently be found by using 
the constraints arising from modelling the data around the caustic passage yielding
the $3+2n$ parameters $(t_\rmn{f}^\star,t_\star^\perp,F^{(s)}_\rmn{r},F^{\star\,(s)}_\rmn{f},
{\hat \omega}_\rmn{f}^\star)$, so that only a 4-dimensional (or 5-dimensional 
if ${\hat \omega}_\rmn{f}^\star$ is ignored) remaining parameter subspace needs to be
searched for solutions.
This search can be performed by creating a suitable parameter grid, and
by assessing the goodness-of-fit
of the arising trial models to the data. For promising regions of the parameter grid,
the grid may be refined in order to succeed towards optimal models and/or the trial models can
be used as initial guesses for fits in the full parameter space.

A convenient parametrization of the remaining subspace is given by $(d,q,\ell,\phi,\zeta_\rmn{f})$,
where the separation parameter $d$ and the mass ratio $q$ are the binary lens characteristics,
$\ell$ stands for the path length along the caustic and yields the location of the fold
singularity ${\vec y}_\rmn{f}$, 
while the caustic crossing angle $\phi$ and the caustic rise parameter
$\zeta_\rmn{f}$ characterize the source trajectory.

For a given fold singularity at ${\vec y}_\rmn{f}$, the local characteristics
$R_\rmn{f}$, ${\vec n}_\rmn{f}$, $A_\rmn{f}$ and $({\vec \nabla} A)_\rmn{f}$ 
are functions of derivatives of the Fermat potential of the lens mapping at the 
critical image ${\vec x}_\rmn{f}$ or the non-critical images ${\vec x}^{(i)}$
which are given by Eqs.~(\ref{eq:calcrf}), (\ref{eq:eigenvec}),
(\ref{eq:calcAf}), and~(\ref{eq:calcnablaAf}), respectively, where ${\vec n}_\rmn{f}$
= ${\vec e}^{(2)}$.
In principle, the binary lens model yields $A_\rmn{other}(t)$ for any source trajectory
and source size. However, as pointed out in Sect.~\ref{sec:magstar}, within 
the validity of the approximation given by  
Eq.~(\ref{eq:expandafoldoth}), $A_\rmn{other}(t) = A^\rmn{p}_\rmn{other}(t)$ 
which itself is simply characterized
by the coefficients $A^\star_\rmn{f}$ and ${\dot A}^\star_\rmn{f}$.
If the value of ${\hat \omega}^\star_\rmn{f}$ is disregarded corresponding 
to a constant $A^\rmn{p}_\rmn{other}$, one obtains
$A^\star_\rmn{f} = A_\rmn{f}$, whereas otherwise
$A^\star_\rmn{f} = A_\rmn{f} - 
\zeta_\rmn{f}\,{\hat \omega}^\star_\rmn{f}\,t_\star^\perp$.

\subsection{Parameter search disregarding ${\hat \omega}_\rmn{f}^\star$}

Let us now investigate how the choice of $(d,q,\ell,\phi,\zeta_\rmn{f})$ together with the
$2+2n$ fold-caustic model parameters $(t_\rmn{f}^\star, t_\star^\perp,
F^{(s)}_\rmn{r}, F^{\star\,(s)}_\rmn{f})$ 
yields the $7+2n$ parameters
($d$, $q$, $u_0$, $\alpha$, $t_0$, $t_{\rm E}$, $\rho_{\star}$, $F_{\rm S}^{(s)}$,
$F_{\rm B}^{(s)}$) which characterize the full lightcurve.

Regardless of the caustic crossing angle $\phi$,
the choice of $\zeta_\rmn{f}$ dictates the values of all
source fluxes $F_\rmn{S}^{(s)}$
and for given $(d,q,\ell)$ the values of
the time-scale of perpendicular motion $t_\rmn{E}^\perp$, the source size parameter
$\rho_\star = \theta_\star/\theta_\rmn{E}$, and all background fluxes
$F_\rmn{B}^{(s)}$.

Explicitly, $t_\rmn{E}^\perp$ is determined by $\zeta_\rmn{f}$ and $R_\rmn{f}$ as
\begin{equation}
t_E^\perp = \frac{\zeta_\rmn{f}^2\,{\hat t}}{R_\rmn{f}}\,,
\end{equation}
while the source size parameter follows from these parameters together with
$t_\star^\perp$ as
\begin{equation}
\rho_\star = \frac{t_\star^\perp}{t_\rmn{E}^\perp} 
= \frac{R_\rmn{f}}{\zeta_\rmn{f}^2\,{\hat t}}\,t_\star^\perp\,.
\end{equation}
Together with the conditions $F^{(m)}_\rmn{S} > 0$ and $A^\star_\rmn{f} > 1$,
Eq.~(\ref{eq:fluxfold}) yields the relations
\begin{eqnarray}
F^{\star(s)}_\rmn{f} - F^{(s)}_\rmn{base} &  = & F^{(s)}_\rmn{S}\,
(A^\star_\rmn{f} - 1) > 0 \,,
\label{eq:flbasesource} \\
F^{\star(s)}_\rmn{f} - F^{(s)}_\rmn{B} &  = & F^{(s)}_\rmn{S}\,
A^\star_\rmn{f} 
\label{eq:flblendsource} > 0\,.
\end{eqnarray}
From these equations and the definition of $F_\rmn{r}^{(s)}$, Eq.~(\ref{eq:defFr}),
the source fluxes $F_\rmn{S}^{(s)}$ and the background fluxes $F_\rmn{B}^{(s)}$
of all $n$ lightcurves follow as
\begin{eqnarray}
F_\rmn{S}^{(s)} & = & \frac{F_\rmn{r}^{(s)}}{\zeta_\rmn{f}}\,, \label{eq:FSzf}\\
F^{(s)}_\rmn{B} & = & F_\rmn{f}^{\star\,{(s)}} - F_\rmn{r}^{(s)}\,
\frac{A_\rmn{f}^\star}{\zeta_\rmn{f}}\,,
\label{eq:FBzf}
\end{eqnarray}
while the baseline fluxes $F^{(s)}_\rmn{base}$ and the blend ratios $g^{(s)}$ are determined as
\begin{eqnarray}
F^{(s)}_\rmn{base} & = & F_\rmn{f}^{\star\,{(s)}} - F_\rmn{r}^{(s)}\,
\frac{A_\rmn{f}^\star-1}{\zeta_\rmn{f}}\,, \\
g^{(s)} & = & \zeta_\rmn{f}\, \frac{F_\rmn{f}^{\star\,{(s)}}}{F_\rmn{r}^{(s)}} -
A_\rmn{f}^\star \label{eq:gzf}\,,
\end{eqnarray}
so that $F^{(s)}_\rmn{S}$, $F^{(s)}_\rmn{B}$, $F^{(s)}_\rmn{base}$ and $g^{(s)}$ depend on no
other parameters than 
$\zeta_\rmn{f}$, $A_\rmn{f}^\star$, 
and the caustic passage flux parameters $F^{(s)}_\rmn{r}$ and 
$F_\rmn{f}^{\star\,{(s)}}$, where $F^{(s)}_\rmn{S}$ even depends on 
$\zeta_\rmn{f}$ and $F^{(s)}_\rmn{r}$ only.

The choice of the caustic crossing angle
$\phi$ determines the
time $t_\star = t_\rmn{E}\,\rho_\star$ in which the source moves by
its angular radius $\theta_\star$
together with the fold-caustic model parameter $t_\star^\perp$ as
\begin{equation}
t_\star = \frac{t_\star^\perp}{\sin \phi} \geq t_\star^\perp\,.
\end{equation}
For given $(d,q,\ell)$ determining ${\vec y}_\rmn{f}$ and
${\vec n}_\rmn{f}$, $\phi$ fixes the
location of the source trajectory determining the parameter $u_0$.

The time-scale of motion $t_\rmn{E} = t_\rmn{E}^\perp\,\sin \phi \leq t_\rmn{E}^\perp$
depends both on $\zeta_\rmn{f}$ and $\phi$, and with $R_\rmn{f}$ reads
\begin{equation}
t_\rmn{E} = \frac{\zeta_\rmn{f}^2\,{\hat t}}{R_\rmn{f}}\,
\sin \phi\,. 
\label{eq:tEmodel}
\end{equation}

Finally, $t_0$ is determined by $t_\rmn{f} = t_\rmn{f}^\star \pm
t_\star^\perp$, the location of the source trajectory given by ${\vec y}_\rmn{f}$,
${\vec n}_\rmn{f}$
and $\phi$, and $t_\rmn{E}$ depending on $\zeta_\rmn{f}$, $\phi$ and $R_\rmn{f}$.

\begin{table*}
\caption{Determination of quantities characterizing the microlensing event
from the fold-caustic parameters ($t_\star^\perp$, $F_\rmn{r}^{(s)}$,
$F_\rmn{f}^{\star\,(s)}$, ${\hat \omega}_\rmn{f}^\star$), the parameters of the 
singularity $(R_\rmn{f}$, ${\vec n}_\rmn{f}$, $A_\rmn{f}$, $({\vec \nabla A})_\rmn{f})$, 
and the adopted caustic crossing angle $\phi$ and caustic rise parameter
$\zeta_\rmn{f}$.}
\label{tab:quantities}
\begin{tabular}{lccccc}
\hline
 & caustic  & & crossing & caustic  & 
 \\ 
\raisebox{1.5ex}[-1.5ex]{parameter}  & passage & \raisebox{1.5ex}[-1.5ex]{singularity} &
angle &
rise & \raisebox{1.5ex}[-1.5ex]{other} \\ \hline
$\mu^\perp = \frac{\theta_\star}{t_\star^\perp}$ & $t_\star^\perp$ & & & & $\theta_\star$ 
\\[0.7ex]
$\mu = 
\frac{\theta_\star}{t_\star^\perp\,\sin \phi}$ & $t_\star^\perp$ & & $\phi$ & & $\theta_\star$ \\
$t_\star = \frac{t_\star^\perp}{\sin \phi}$ & $t_\star^\perp$ & & $\phi$ & & \\[0.7ex]
$\rho_\star 
= \frac{R_\rmn{f}}{\zeta_\rmn{f}^2\,{\hat t}}\,t_\star^\perp$
& $t_\star^\perp$ & $R_\rmn{f}$ & & $\zeta_\rmn{f}$ & 
\\[0.7ex]
$t_E^\perp = \frac{\zeta_\rmn{f}^2\,{\hat t}}{R_\rmn{f}}$ & & $R_\rmn{f}$ &
& $\zeta_\rmn{f}$ & \\[0.7ex]
$t_\rmn{E} = \frac{\zeta_\rmn{f}^2\,{\hat t}}{R_\rmn{f}}\,
\sin \phi$ & & $R_\rmn{f}$ &
$\phi$ & $\zeta_\rmn{f}$ &\\[0.7ex]
$F_\rmn{S}^{(s)} = \frac{F_\rmn{r}^{(s)}}{\zeta_\rmn{f}}$ & $F_\rmn{r}^{(s)}$ &
& &  $\zeta_\rmn{f}$ & \\[0.7ex]
$F^{(s)}_\rmn{B} = F_\rmn{f}^{\star\,{(s)}} - F_\rmn{r}^{(s)}\,
\frac{A_\rmn{f}^\star}{\zeta_\rmn{f}}$ & $F_\rmn{r}^{(s)}$, $F_\rmn{f}^{\star\,{(s)}}$ &
$A^\star_\rmn{f}$ & &  $\zeta_\rmn{f}$ & 
\\
\hline
\end{tabular}

\medskip
The choice of a binary lens $(d,q)$ and of the singularity at ${\vec y}_\rmn{f}$ characterized
by $\ell$ yields $R_\rmn{f}$, ${\vec n}_\rmn{f}$, $A_\rmn{f}$, 
and $({\vec \nabla} A)_\rmn{f}$.
If one disregards ${\hat \omega}_\rmn{f}^\star$, $A^\star_\rmn{f} = A_\rmn{f}$
and the value of $({\vec \nabla} A)_\rmn{f}$ is redundant, whereas otherwise
$A^\star_\rmn{f} = A_\rmn{f}-\zeta_\rmn{f}\,{\hat \omega}_\rmn{f}^\star\,
t_\star^\perp$ depends on $\zeta_\rmn{f}$ and ${\hat \omega}_\rmn{f}^\star$,
and ${\hat \omega}_\rmn{f}^\star$, $R_\rmn{f}$, ${\vec n}_\rmn{f}$ 
and $({\vec \nabla} A)_\rmn{f}$ 
constrain $\phi$ and provide a relation to $\zeta_\rmn{f}$.
\end{table*}

Table~\ref{tab:quantities} summarizes the dependencies of the  
quantities characterizing the microlensing event,
namely the proper motion $\mu$ and the time-scale of 
motion $t_\rmn{E}$ as well as their
transverse components $\mu^\perp$ and $t_\rmn{E}^\perp$,
the time $t_\star$ in which the sources moves by its angular radius
$\theta_\star$
relative to the lens,
the source size parameter $\rho_\star = \theta_\star/\theta_\rmn{E} = 
t_\star/t_\rmn{E}$, 
and the source and
background fluxes $F_\rmn{S}^{(s)}$ and
$F_\rmn{B}^{(s)}$, 
on the fold-caustic parameters ($t_\star^\perp$, $F_\rmn{r}^{(s)}$,
$F_\rmn{f}^{\star\,(s)}$, ${\hat \omega}_\rmn{f}^\star$), the parameters of the 
singularity $(R_\rmn{f}$, ${\vec n}_\rmn{f}$, $A_\rmn{f}$, $({\vec \nabla A})_\rmn{f})$, 
and the adopted caustic crossing angle $\phi$ and caustic rise parameter
$\zeta_\rmn{f}$.

\begin{figure}
\includegraphics[width=84mm]{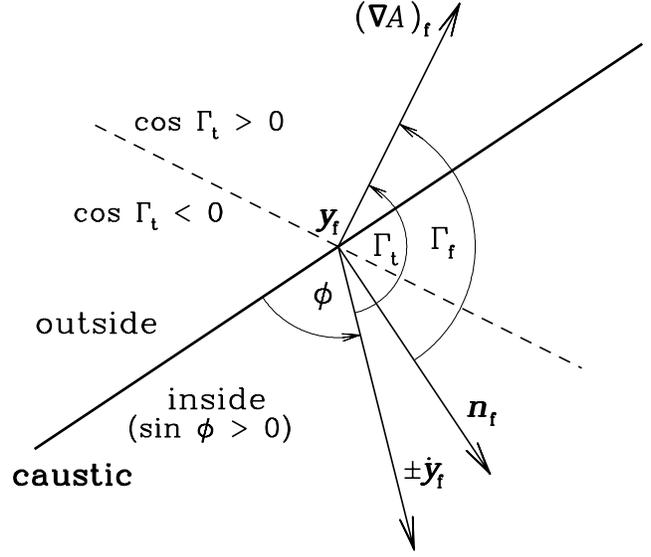}
\caption{Definition of the angles $\phi$, $\Gamma_\rmn{f}$, and
$\Gamma_\rmn{t}$ which describe the relative orientation between the
source trajectory characterized by $\pm\,\dot{\vec{y}}_\rmn{f}$, the
caustic characterized by its inside-pointing normal ${\vec n}_\rmn{f}$, 
and the gradient of the magnification $({\vec\nabla} A)_\rmn{f}$ 
of the non-critical images at the fold singularity 
${\vec y}_\rmn{f}$. Regions are marked for which the source trajectory 
matches the indicated
conditions: $\sin \phi > 0$ ($0 < \phi < \upi$) follows from the requirement that
$\pm\,\dot{\vec{y}}_\rmn{f}$ points inside, while the condition 
 $\zeta_\rmn{f} > 0$ requires either
 $\cos \Gamma_\rmn{t} >0$ for ${\hat \omega}_\rmn{f}^\star
 > 0$,   $\cos \Gamma_\rmn{t} <0$ for ${\hat \omega}_\rmn{f}^\star
 < 0$,  or  $\cos \Gamma_\rmn{t} = 0$ for ${\hat \omega}_\rmn{f}^\star
 = 0$.}
\label{fig:angles}
\end{figure}

\subsection{Parameter search using ${\hat \omega}_\rmn{f}^\star$}
If the complete set of $3+2n$ fold-caustic model parameters 
$(t_\rmn{f}^\star, t_\star^\perp, F^{(s)}_\rmn{r}, F^{\star\,(s)}_\rmn{f},
{\hat \omega}_\rmn{f}^\star)$ is used in the search of the remaining parameters 
required to characterize the full
lightcurve, the parameter ${\hat \omega}_\rmn{f}^\star$ provides a relation between the
caustic crossing angle $\phi$ and the caustic rise time $\zeta_\rmn{f}$, so that the
parameter search is reduced to the four-dimensional subspaces $(d,q,\ell,\phi)$ or
$(d,q,\ell,\zeta_\rmn{f})$.

As an alternative to expanding
$A^\rmn{p}_\rmn{other}(t)$ which is assumed to vary slowly during
the progress of the caustic passage around $t_\rmn{f}^\star$,
it may also be approximated by a linear expansion around 
$t_\rmn{f}$ which reads 
\begin{equation}
A^\rmn{p}_\rmn{other}(t)
 \simeq 
A_\rmn{f}+  (t-t_\rmn{f}) \dot{A}_\rmn{f}\,,
\label{eq:expandafoldothtf}
\end{equation}
where $A_\rmn{f} = 
A^\rmn{p}_\rmn{other}(t_\rmn{f})$ (as defined in Sect.~\ref{sec:magstar}) and
$\dot{A}_\rmn{f} = 
\dot{A}^\rmn{p}_\rmn{other}(t_\rmn{f})$, while
the corresponding expansion of $A^\rmn{p}_\rmn{other}({\vec y}(t))$ in space 
around ${\vec y}_\rmn{f}$ 
reads (c.f.\ Eq.~(\ref{eq:Aotherexpand}))
\begin{equation}
A^\rmn{p}_\rmn{other}({\vec y}(t))
\simeq  A_\rmn{f}
+ (t-t_\rmn{f})\,\dot{\vec{y}}_\rmn{f}\cdot
({\vec\nabla} A)_\rmn{f}
\,.
\label{eq:expandafoldothfsp}
\end{equation}
The comparison of Eq.~(\ref{eq:expandafoldothtf}) with the expansion around $t^\star_\rmn{f}$,
Eq.~(\ref{eq:expandafoldoth}), yields the relations 
$\dot{A}^\star_\rmn{f} = \dot{A}_\rmn{f}$ and
$A^\star_\rmn{f} = A_\rmn{f} \mp \dot{A}^\star_\rmn{f}\,t_\star^\perp$, and by comparing
Eq.~(\ref{eq:expandafoldothfsp}) with Eq.~(\ref{eq:expandafoldothtf}), one obtains
\begin{eqnarray}
\dot{A}^\star_\rmn{f} & = & \dot{\vec{y}}_\rmn{f} \cdot 
({\vec\nabla} A)_\rmn{f} 
\nonumber \\
& = & \pm\,|\dot{\vec{y}}_\rmn{f}|\,|({\vec\nabla} A)_\rmn{f}|\,
\cos \Gamma_\rmn{t}\,,
\label{eq:adot1}
\end{eqnarray}
where $\Gamma_\rmn{t}$ denotes the angle measured from 
$\pm\,\dot{\vec{y}}_\rmn{f}$ to 
$({\vec{\nabla}} A)_\rmn{f}$.
As a function of the angle $0 \leq \Gamma_\rmn{f} < 2\upi$ measured from the caustic normal 
${\vec n}_\rmn{f}$ 
to $({\vec\nabla} A)_\rmn{f}$, which is determined by the properties of the lens mapping at
${\vec y}_\rmn{f}$ alone, and the caustic crossing angle 
$0 < \phi < \upi$,
$\Gamma_\rmn{t}$ reads
\begin{equation}
\Gamma_\rmn{t} = \Gamma_\rmn{f}-\phi+\frac{\upi}{2}
-2\,\upi\,\left\lfloor\frac{\Gamma_\rmn{f}-\phi}{2\,\upi}+\frac{1}{2}\right\rfloor\,,
\label{eq:gammt}
\end{equation}
forcing $-\frac{\upi}{2} \leq \Gamma_\rmn{t} < \frac{3\,\upi}{2}$.
With this definition, $0 < \phi < \upi$ increases with a counterclockwise
rotation of $\pm\,\dot{\vec{y}}_\rmn{f}$.
The definition of the angles $\phi$, $\Gamma_\rmn{f}$, and
$\Gamma_\rmn{t}$ is illustrated in Fig.~\ref{fig:angles}.

\begin{table*}
\caption{Constraints on the orientation of the source trajectory imposed by the sign
of ${\hat \omega}_\rmn{f}^\star$ depending on the orientation of
$(\vec \nabla A)_\rmn{f}$ with respect to the caustic.}
\label{tab:restrictangle}
\begin{tabular}{cccc}
\hline
${\hat \omega}_\rmn{f}^\star >0$ & $\Gamma_\rmn{f} \in (0,\upi)$ &
$-\frac{\upi}{2} < \Gamma_\rmn{f}-\frac{\upi}{2} < \Gamma_\rmn{t} < \frac{\upi}{2}$ &
$0 < \Gamma_\rmn{f} < \phi < \upi$ \\[0.18ex]
${\hat \omega}_\rmn{f}^\star >0$ & $\Gamma_\rmn{f} \in (\upi,2\upi)$ &
$-\frac{\upi}{2} < \Gamma_\rmn{t} < \Gamma_\rmn{f}-\frac{3\upi}{2} < \frac{\upi}{2}$ & 
$0 < \phi < \Gamma_\rmn{f}-\upi <  \upi$ \\[0.18ex]
${\hat \omega}_\rmn{f}^\star >0$ & $\Gamma_\rmn{f} = 0$ &
$-\frac{\upi}{2} < \Gamma_\rmn{t} <  \frac{\upi}{2}$ & 
$0 < \phi < \upi$ \\[0.18ex]
${\hat \omega}_\rmn{f}^\star >0$ & $\Gamma_\rmn{f} = \upi$ &
\multicolumn{2}{c}{\em no viable trajectory} 
\\[0.18ex]
${\hat \omega}_\rmn{f}^\star <0$ & $\Gamma_\rmn{f} \in (0,\upi)$ &
$\frac{\upi}{2} < \Gamma_\rmn{t} < \Gamma_\rmn{f}+\frac{\upi}{2} < \frac{3\upi}{2}$ & 
$0 < \phi < \Gamma_\rmn{f} <  \upi$ \\[0.18ex]
${\hat \omega}_\rmn{f}^\star <0$ & $\Gamma_\rmn{f} \in (\upi,2\upi)$ &
$\frac{\upi}{2} < \Gamma_\rmn{f}-\frac{\upi}{2} < \Gamma_\rmn{t} < \frac{3\upi}{2}$ & 
$0 < \Gamma_\rmn{f}-\upi < \phi <  \upi$ \\[0.18ex]
${\hat \omega}_\rmn{f}^\star <0$ & $\Gamma_\rmn{f} = 0$ &
\multicolumn{2}{c}{\em no viable trajectory} 
\\[0.18ex]
${\hat \omega}_\rmn{f}^\star <0$ & $\Gamma_\rmn{f} = \upi$ &
$\frac{\upi}{2} < \Gamma_\rmn{t} <  \frac{3\upi}{2}$ & 
$0 < \phi < \upi$ \\[0.18ex]
${\hat \omega}_\rmn{f}^\star = 0$ & $\Gamma_\rmn{f} \in (0,\upi)$ &
$\Gamma_\rmn{t} = \frac{\upi}{2}$ & 
$\phi = \Gamma_\rmn{f}$ \\[0.18ex]
${\hat \omega}_\rmn{f}^\star = 0$ & $\Gamma_\rmn{f} \in (\upi,2\upi)$ &
$\Gamma_\rmn{t} = -\frac{\upi}{2}$ & 
$\phi = \Gamma_\rmn{f} -\upi$ \\[0.18ex]
${\hat \omega}_\rmn{f}^\star = 0$ & $\Gamma_\rmn{f} = 0$ or $\Gamma_\rmn{f} = \upi$&
\multicolumn{2}{c}{\em no viable trajectory} 
\\[0.18ex]

\hline
\end{tabular}

\medskip
The fold-caustic model parameter ${\hat \omega}_\rmn{f}^\star$ 
characterizes the temporal variation of the
magnification due to non-critical images, $0 \leq \Gamma_\rmn{f} = \angle(\vec n_\rmn{f},
(\vec \nabla A)_\rmn{f})< 2\upi$ denotes
the orientation of its gradient $(\vec \nabla A)_\rmn{f}$ relative to the caustic inside normal
$\vec n_\rmn{f}$
at the fold singularity $\vec y_\rmn{f}$, 
$\Gamma_\rmn{t} =\angle(\pm \dot{\vec{y}}_\rmn{f},(\vec \nabla A)_\rmn{f})$ is the angle 
between the inside pointing tangent to the source trajectory $\pm \dot{\vec{y}}_\rmn{f}$ 
 and $(\vec \nabla A)_\rmn{f}$, and $\phi$ denotes the
caustic crossing angle.
\end{table*}

With $\zeta_\rmn{f} > 0$ and 
$\dot{A}^\star_\rmn{f} = \pm\,\zeta_\rmn{f}\,{\hat \omega}^\star_\rmn{f}$, Eq.~(\ref{eq:adot1})
requires the sign of $\cos \Gamma_\rmn{t}$ and ${\hat \omega}^\star_\rmn{f}$ to coincide.
With Eq.~(\ref{eq:gammt}) and the condition $0 < \phi < \upi$, this results
in the allowed ranges for $\Gamma_\rmn{t}$ and $\phi$ shown in 
Table~\ref{tab:restrictangle}. 
Configurations $(d,q,\ell)$ for which there is no viable trajectory have to be 
discarded.

By inserting
\begin{equation}
\cos \Gamma_\rmn{t} = \cos \Gamma_\rmn{f}\,
\sin\phi - \sin \Gamma_\rmn{f}\,\cos\phi\,,
\label{eq:cosGammat}
\end{equation}
and $|\dot{\vec{y}}_\rmn{f}| = 1/t_\rmn{E}$ as given by Eq.~(\ref{eq:tEmodel})
into Eq.~(\ref{eq:adot1}), and using ${\hat \omega}^\star_\rmn{f} = \pm 
\dot{A}^\star_\rmn{f}/\zeta_\rmn{f}$, one obtains
\begin{equation}
{\hat \omega}^\star_\rmn{f} = \frac{R_\rmn{f}\,|({\vec\nabla} A)_\rmn{f}|}
{\zeta_\rmn{f}^3\,\hat t}\,
K_\rmn{f}(\phi)
\,,
\label{eq:omegaphizeta}
\end{equation}
where
\begin{equation}
K_\rmn{f}(\phi) = \cos \Gamma_\rmn{f} - \sin \Gamma_\rmn{f}\,\cot \phi\,.
\end{equation}
For ${\hat \omega}^\star_\rmn{f} \neq 0$ and $\sin \Gamma_\rmn{f} \neq 0$, this
yields a relation between $\phi$ and $\zeta_\rmn{f}$, and, according to the previous
subsection also between the time-scale $t_\rmn{E}$ and the source and background fluxes
$F_\rmn{S}^{(s)}$ and $F_\rmn{B}^{(s)}$, respectively.
With
\begin{equation}
\zeta_{\rmn{f,ref}} = \left(\frac{R_\rmn{f}\,|({\vec\nabla} A)_\rmn{f}|}
{{\hat \omega}^\star_\rmn{f}\,\hat t}\right)^{1/3}\,,
\label{eq:zetaf0}
\end{equation}
one finds explicitly
\begin{equation}
\zeta_\rmn{f}(\phi) = \zeta_{\rmn{f,ref}}\,\left[K_\rmn{f}(\phi)\right]^{1/3}
\label{eq:zetaofphi}
\end{equation}
or
\begin{equation}
\phi(\zeta_\rmn{f}) = \frac{\upi}{2} + \arctan \frac{(\zeta_\rmn{f}/\zeta_\rmn{f,ref})^3 - \cos \Gamma_\rmn{f}}
{\sin \Gamma_\rmn{f}}\,.
\label{eq:phiofzeta}
\end{equation}
For $\Gamma_\rmn{f} = 0$, $\zeta_\rmn{f}$ adopts the
value $\zeta_{\rmn{f,ref}}$ (while ${\hat \omega}^\star_\rmn{f} >0$),
whereas $-\zeta_{\rmn{f,ref}}$ is adopted for $\Gamma_\rmn{f} = \upi$
 (while ${\hat \omega}^\star_\rmn{f} <0$), irrespective
of $\phi$.
For ${\hat \omega}^\star_\rmn{f} = 0$, one finds that
$\phi = \Gamma_\rmn{f}$ for $\Gamma_\rmn{f} \in (0,\upi)$ or
$\phi = \Gamma_\rmn{f}-\upi$ for $\Gamma_\rmn{f} \in (\upi,2\upi)$.
Therefore, the remaining parameter space can be parametrized by 
$(d, q, \ell, \phi)$ for ${\hat \omega}^\star_\rmn{f} \neq 0$ and by
$(d, q, \ell, \zeta_\rmn{f})$ for $\sin \Gamma_\rmn{f} \neq 0$, whereas otherwise such
a parametrization is not viable due to the adoption of a single fixed value.

In analogy to $\zeta_\rmn{f}$, one obtains $\dot{A}^\star_\rmn{f} = \pm 
\zeta_\rmn{f} {\hat \omega}^\star_\rmn{f}$ as function of the caustic crossing angle $\phi$ as
\begin{equation}
\dot{A}^\star_\rmn{f}(\phi) = \dot{A}^\star_{\rmn{f,ref}}\,\left[K_\rmn{f}(\phi)\right]^{1/3}\,,
\label{eq:astarfdot}
\end{equation}
where
\begin{equation}
\dot{A}^\star_{\rmn{f,ref}} = \pm \left(
\frac{({\hat \omega}^\star_\rmn{f})^2\,R_\rmn{f}\,
|({\vec\nabla} A)_\rmn{f}|}{{\hat t}}\right)^{1/3}\,.
\label{eq:afp0}
\end{equation}

\begin{figure}
\includegraphics[width=84mm]{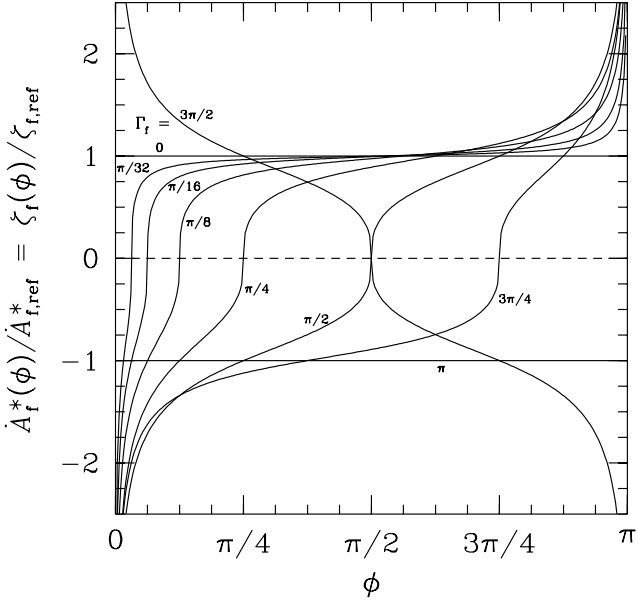}
\caption{Variation of the magnification of the non-critical images $\dot{A}^\star_\rmn{f}$ 
as well as the caustic rise parameter $\zeta_\rmn{f}$ as
a function of the caustic crossing angle $\phi$ for selected values of the angle $\Gamma_\rmn{f}$ 
between caustic normal ${\vec n}_\rmn{f}$ and gradient of the magnification 
$({\vec\nabla} A)_\rmn{f}$ of the non-critical
images at the fold singularity ${\vec y}_\rmn{f}$. 
With $\dot{A}^\star_{\rmn{f,ref}}$ given by Eq.~(\ref{eq:afp0}) and
$\zeta_{\rmn{f,ref}}$ given by Eq.~(\ref{eq:zetaf0}),
the curves show
$\dot{A}^\star_\rmn{f}(\phi)/\dot{A}^\star_{\rmn{f,ref}}
= \zeta_\rmn{f}/\zeta_{\rmn{f,ref}} = [K_\rmn{f}(\phi)]^{1/3}$
for $\Gamma_\rmn{f} = 0$, $\upi/32$, $\upi/16$,
$\upi/8$, $\upi/4$, $\upi/2$, $3\upi/4$, $\upi$, and~$3 \upi/2$.
The range for the crossing angle $\phi$ is restricted as shown in Table~\ref{tab:restrictangle},
so that $\zeta_\rmn{f}(\phi) > 0$ and the sign of $\dot{A}^\star_\rmn{f}(\phi)$
matches that of ${\hat \omega}^\star_\rmn{f}$.}
\label{fig:afp}
\end{figure}

The function $[K_\rmn{f}(\phi)]^{1/3} = \zeta_\rmn{f}(\phi)/\zeta_\rmn{f,ref}
= \dot{A}^\star_\rmn{f}(\phi)/\dot{A}^\star_\rmn{f,ref}$ is shown in Fig.~\ref{fig:afp} for
selected angles $\Gamma_\rmn{f}$.
$[K_\rmn{f}(\phi)]^{1/3}$ is strictly  monotonic
in $\phi$ except for $\Gamma_\rmn{f} = 0$ and $\Gamma_\rmn{f} = \upi$, where 
the constant value $\pm 1$ is adopted.
With the requirement of $\zeta_\rmn{f}(\phi) > 0$ and the sign of $\dot{A}^\star_\rmn{f}(\phi)$
to match that of ${\hat \omega}^\star_\rmn{f}$, the range of $\phi$ is restricted as
shown in Table~\ref{tab:restrictangle}.
For $\phi \to 0$ or
$\phi \to \upi$, $|[K_\rmn{f}(\phi)]^{1/3}|$ tends to infinity,
whereas zero is approached for $\phi \to \Gamma_\rmn{f}$ or $\phi \to \Gamma_\rmn{f}- \upi$,
respectively.

\begin{figure}
\includegraphics[width=84mm]{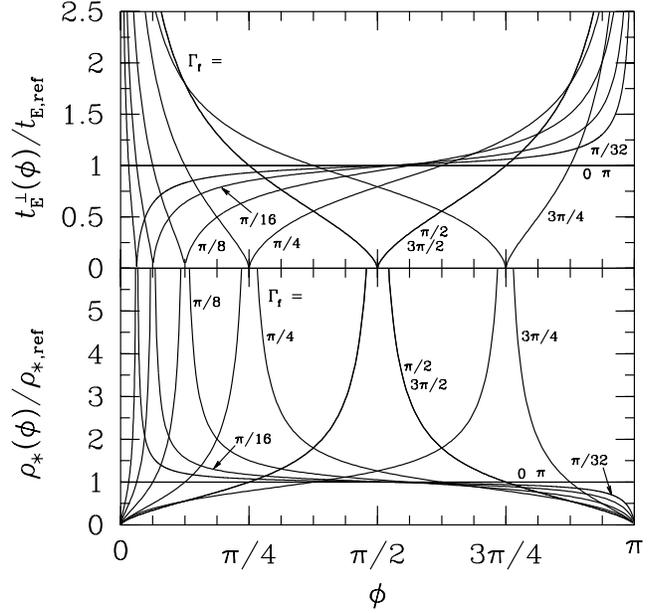}
\caption{The dependence of the time-scale of transverse motion $t_\rmn{E}^\perp$ and the
source size parameter $\rho_\star = \theta_\star/\theta_\rmn{E} = t_\star^\perp/t_\rmn{E}^\perp$ 
on the caustic crossing angle $\phi$. The upper panel shows curves for 
$t_\rmn{E}^\perp(\phi)/t_{\rmn{E,ref}} = [K_\rmn{f}(\phi)]^{2/3}$ 
while the lower panel shows the corresponding curves
for 
$\rho_\star(\phi)/\rho_{\star,\rmn{ref}} = [K_\rmn{f}(\phi)]^{-2/3}$ for 
the angle 
between caustic normal ${\vec n}_\rmn{f}$ and gradient of the magnification 
of the non-critical
images at the fold singularity ${\vec y}_\rmn{f}$ assuming the values 
$\Gamma_\rmn{f} = 0$, $\upi/32$, $\upi/16$,
$\upi/8$, $\upi/4$, $\upi/2$, $3\upi/4$, $\upi$, and~$3 \upi/2$. The reference values
$t_{\rmn{E,ref}}$ and $\rho_{\star,\rmn{ref}}$ are given by 
Eqs.~(\ref{eq:tEref}) and~(\ref{eq:rhostarref}), respectively.
According to the sign of ${\hat \omega}^\star_\rmn{f}$,
the caustic crossing angle $\phi$ is either restricted to the range
$\Gamma_\rmn{f} < \phi < \upi$ or $0 < \phi < \Gamma_\rmn{f}$
 for $\Gamma_\rmn{f} \in (0,\upi)$
or to $\Gamma_\rmn{f}-\upi < \phi < \upi$ or $0 < \phi < \Gamma_\rmn{f}-\upi$ 
for $\Gamma_\rmn{f} \in (\upi,2\upi)$ as shown in
Table~\ref{tab:restrictangle}, so that $[K_\rmn{f}(\phi)]^{2/3}$  and 
$[K_\rmn{f}(\phi)]^{-2/3}$ are strictly monotonic within the allowed range except for
$\Gamma_\rmn{f} = 0$ or $\Gamma_\rmn{f} = \upi$, where a constant value is adopted.} 
\label{fig:tpr}
\end{figure}

With the caustic rise parameter $\zeta_\rmn{f}$ depending 
on the caustic crossing angle $\phi$ as described by Eq.~(\ref{eq:zetaofphi}), 
the time-scale of transverse motion $t_\rmn{E}^\perp$ and the source size parameter 
$\rho_\star = t_\star^\perp/t_\rmn{E}^\perp$ also
become functions of $\phi$ which read
\begin{equation}
t_\rmn{E}^\perp(\phi) = t_\rmn{E,ref}\,\left[K_\rmn{f}(\phi)\right]^{2/3}
\end{equation}
and
\begin{equation}
\rho_\star(\phi) = \rho_{\star,\rmn{ref}}\,\left[K_\rmn{f}(\phi)\right]^{-2/3}\,,
\end{equation}
respectively, where
\begin{equation}
t_\rmn{E,ref} = 
R_\rmn{f}^{-1/3}\,\left(\frac{|({\vec\nabla} A)_\rmn{f}|}
{{\hat \omega}^\star_\rmn{f}\,\hat t}\right)^{2/3}\,{\hat t}\,
\label{eq:tEref}
\end{equation}
and
\begin{equation}
\rho_{\star,\rmn{ref}} = 
R_\rmn{f}^{1/3}\,\left(\frac{{\hat \omega}^\star_\rmn{f}\,\hat t}{|({\vec\nabla} A)_\rmn{f}|}
\right)^{2/3}\,\frac{t_\star^\perp}{\hat t}\,.
\label{eq:rhostarref}
\end{equation}
For $\Gamma_\rmn{f} = 0$ or $\Gamma_\rmn{f} = \upi$ (i.e.~$\sin \Gamma_\rmn{f} = 0$),
$t_\rmn{E}^\perp = t_\rmn{E,ref}^\perp$ and $\rho_\star = \rho_{\star,\rmn{ref}}$ 
for any $\phi$, whereas otherwise $t_\rmn{E}^\perp(\phi)$ and $\rho_\star(\phi)$ 
are strictly monotonic within the allowed parameter range for $\phi$ shown
in Table~\ref{tab:restrictangle},
where $t_\rmn{E}^\perp$ reaches zero and $\rho_\star$ tends to infinity for
$\phi \to \Gamma_\rmn{f}$ or $\phi \to \Gamma_\rmn{f}-\upi$, while
$t_\rmn{E}^\perp$ tends to infinity and $\rho_\star$ approaches zero for
 $\phi \to 0$ or $\phi \to \upi$.

\begin{figure}
\includegraphics[width=84mm]{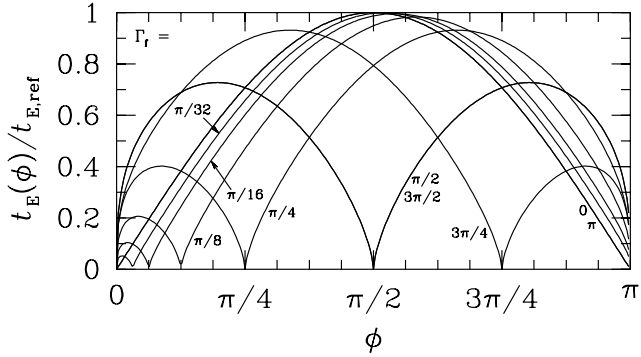}
\caption{The time-scale of motion $t_\rmn{E}$ as a function
of  
the caustic crossing angle $\phi$ for different selected angles $\Gamma_\rmn{f}$ 
between caustic normal ${\vec n}_\rmn{f}$ and gradient of the magnification 
$({\vec\nabla} A)_\rmn{f}$ of the non-critical
images at the fold singularity  ${\vec y}_\rmn{f}$. 
The curves show
$t_\rmn{E}(\phi)/t_{\rmn{E,ref}}
= [K_\rmn{f}(\phi)]^{2/3}\,\sin \phi$, where $t_\rmn{E,ref}$ is given by Eq.~(\ref{eq:tEref}),
for $\Gamma_\rmn{f} = 0$, $\upi/32$, $\upi/16$,
$\upi/8$, $\upi/4$, $\upi/2$, $3\upi/4$, $\upi$, and~$3 \upi/2$.
According to Table~\ref{tab:restrictangle}, the condition $\zeta_\rmn{f} > 0$ 
restricts $\phi$ either to the range
$\Gamma_\rmn{f} < \phi < \upi$ or $0 < \phi < \Gamma_\rmn{f}$
 for $\Gamma_\rmn{f} \in (0,\upi)$
or to $\Gamma_\rmn{f}-\upi < \phi < \upi$ or $0 < \phi < \Gamma_\rmn{f}-\upi$
for $\Gamma_\rmn{f} \in (\upi,2\upi)$
depending on the sign of ${\hat \omega}^\star_\rmn{f}$. Therefore, 
$t_\rmn{E}(\phi)$ has a unique maximum within the allowed range of caustic crossing angles
$\phi$, whereas $t_\rmn{E}(\phi)$ approaches zero at its boundaries.}
\label{fig:tephi}
\end{figure}

The time-scale of motion $t_\rmn{E}$ can also be written  
as function of the caustic crossing angle $\phi$ reading 
\begin{equation}
t_\rmn{E}(\phi) = t_\rmn{E,ref}\,\left[K_\rmn{f}(\phi)\right]^{2/3}\,\sin \phi\,,
\end{equation}
with $t_\rmn{E,ref}$ as defined by Eq.~(\ref{eq:tEref}),
so that $t_\rmn{E}$ reaches a unique maximum within the allowed range for
$\phi$ and approaches zero at its boundaries, located
at $\phi = 0$, $\phi = \upi$, $\phi = \Gamma_\rmn{f}$ or $\phi = \Gamma_\rmn{f}-\upi$.
Antiparallel 
orientations of the  
gradient 
of the magnification of the non-critical images relative to the caustic normal for which 
$\Gamma_\rmn{f}$ differs by (an odd multiple of) $\upi$ yield identical values for 
$t_\rmn{E}^\perp$, $\rho_\star$, and $t_\rmn{E}$.   
Fig.~\ref{fig:tpr} shows $t_\rmn{E}^\perp(\phi)/t_\rmn{E,ref} =
 [K_\rmn{f}(\phi)]^{2/3}$ and
$\rho_\star(\phi)/\rho_{\star,\rmn{ref}} = [K_\rmn{f}(\phi)]^{-2/3}$ 
while Fig.~\ref{fig:tephi} shows
$t_\rmn{E}(\phi)/t_\rmn{E,ref} = [K_\rmn{f}(\phi)]^{2/3}|\,\sin \phi$ for the same 
selected values of $\Gamma_\rmn{f}$ as chosen for Fig.~\ref{fig:afp}.

Depending on the sign of ${\hat \omega}_\rmn{f}^\star$, 
$A^\rmn{p}_\rmn{other}(t)$ as given by the
expansion 
in Eq.~(\ref{eq:expandafoldoth})
yields unphysical values $A^\rmn{p}_\rmn{other}(t) \leq 1$ for either sufficiently 
large or small $t$. For the region containing data near the caustic passage and at least during
the caustic passage itself, one should require $A^\rmn{p}_\rmn{other}(t) > 1$. This condition
yields an upper limit for the caustic rise parameter $\zeta_\rmn{f}$, which is limited
to the range
\begin{equation}
\zeta_\rmn{f} < \zeta_\rmn{f,var} = \frac{A_\rmn{f}-1}{|{\hat \omega}_\rmn{f}^\star|\,\tau}
\end{equation}
where $\tau \geq t_\star^\perp$ denotes the timespan from 
$t_\rmn{f}$ toward the outside of the caustic 
for ${\hat \omega}_\rmn{f}^\star > 0$ or toward the inside of the caustic 
for ${\hat \omega}_\rmn{f}^\star < 0$, 
so that the constraint arises from fulfulling
$A^\rmn{p}_\rmn{other}(t) > 1$ at the  
time $t_\rmn{f}-\tau$ for $\pm\,{\hat \omega}_\rmn{f}^\star > 0$ 
and $t_\rmn{f}+\tau$ for $\pm\,{\hat \omega}_\rmn{f}^\star < 0$, while there is no constraint
for ${\hat \omega}_\rmn{f}^\star = 0$. 
By means of Eq.~(\ref{eq:phiofzeta}), the
constraint on $\zeta_\rmn{f}$
replaces the boundary at $\phi = 0$ or $\phi = \pi$
with $\phi_\rmn{var} = \phi(\zeta_\rmn{f})$, where the allowed range
depends on the sign of
${\hat \omega}_\rmn{f}^\star$ and the orientation of $(\vec \nabla A)_\rmn{f}$ with respect
to the caustic as shown in Table~\ref{tab:restrictanglevarblend}.
For $\Gamma_\rmn{f} = 0$ or $\Gamma_\rmn{f} = \upi$, $\zeta_\rmn{f}$ adopts the fixed
value $|\zeta_\rmn{f,ref}|$ irrespective of the caustic
crossing angle $\phi$, so that no restriction on $\phi$ results if
$|\zeta_\rmn{f,ref}| < \zeta_\rmn{f,var}$, whereas otherwise the chosen configuration
characterized by $(d, q, \ell, \zeta_\rmn{f})$ has to be
discarded.

\subsection{Blending constraints}
In the absence of blending, i.e.  $F_\rmn{B}^{(s)} = 0$ or $g^{(s)} = 0$ for all
lightcurves, Eq.~(\ref{eq:defgAr}) fixes $\zeta_\rmn{f}$ to
\begin{equation}
\zeta_\rmn{f} = \frac{A^\star_\rmn{f}}{g_{\rmn{f}}^\star}\,,
\end{equation}
where $g_{\rmn{f}}^\star$ is a fold-caustic model parameter. 
If one disregards ${\hat \omega}_\rmn{f}^\star$, the parameter search is reduced to the
four-dimensional subspace
$(d,q,\ell,\phi)$. Otherwise, $A^\star_\rmn{f} = A_\rmn{f} - \zeta_\rmn{f}\,
{\hat \omega}_\rmn{f}^\star\,t_\star^\perp$, so that
\begin{equation}
\zeta_\rmn{f} = \frac{A_\rmn{f}}{g_\rmn{f}^\star+
{\hat \omega}_\rmn{f}^\star\,t_\star^\perp}\,,
\end{equation}
and $\phi$ is determined by means of Eq.~(\ref{eq:phiofzeta}), 
leaving only the three-dimensional subspace $(d,q,\ell)$ to be explored.

Restricting the parameter space to non-negative background fluxes,
i.e.~$F_\rmn{B}^{(s)} \geq 0$
or $g^{(s)} \geq  0$, yields a lower limit to $\zeta_\rmn{f}$, limiting it to
the range
\begin{equation}
\zeta_\rmn{f} \geq \zeta_\rmn{f,B} = \frac{A_\rmn{f}}
{\hat g_{\rmn{f},\rmn{min}}^\star
+{\hat \omega}_\rmn{f}^\star\,t_\star^\perp}\,,
\label{eq:zetablendlimit}
\end{equation}
with $g_{\rmn{f},\rmn{min}}^\star > 0$ being defined by Eq.~(\ref{eq:defgmin}).
The condition $g_{\rmn{f},\rmn{min}}^\star > 0$ corresponds to  
the fluxes for all sites 
at time $t_\rmn{f}^\star$ (when the leading limb of the source enters or its
 trailing limb exits the
caustic) to fulfill $F_\rmn{f}^{\star\,(s)} > 0$ for 
$F_\rmn{B}^{(s)} \geq 0$, whereas at least one of the
conditions $F_\rmn{S}^{(s)} > 0$ and $A^\star_\rmn{f} > 1$ would be violated
if $F_\rmn{f}^{\star\,(s)} \leq 0$.
Similarly, the simulataneous fulfillment
of $F_\rmn{S}^{(s)} > 0$, $A_\rmn{f} > 1$ and
$F_\rmn{B}^{(s)}$ implies the fluxes at time $t_\rmn{f}$ (when the source
center crosses the caustic) to be restricted
to 
$F_\rmn{fold}(t_\rmn{f}) = F_\rmn{f}^{\star\,(s)} + F_\rmn{r}^{(s)}\,
{\hat \omega}_\rmn{f}^\star\,t_\star^\perp$  
$= F_\rmn{S}^{(s)} A_\rmn{f} + F_\rmn{B}^{(s)} > 0$, so that
$g_{\rmn{f},\rmn{min}}^\star
+{\hat \omega}_\rmn{f}^\star\,t_\star^\perp > 0$  needs to be fulfilled,
posing a further restriction in the case ${\hat \omega}_\rmn{f}^\star < 0$.

The blending limit on $\zeta_\rmn{f}$ given by Eq.~(\ref{eq:zetablendlimit}) 
implies a lower limit to the time-scale of perpendicular motion
\begin{equation}
t_\rmn{E}^\perp \geq \frac{\zeta_\rmn{f,B}^2}{R_\rmn{f}}\,{\hat t}\,,
\end{equation}
and an upper limit to the source size parameter
\begin{equation}
\rho_\star \leq \frac{R_\rmn{f}}{\zeta_\rmn{f,B}^2}\,
\frac{t_\star^\perp}{\hat t}\,,
\end{equation}
while the ranges of the fluxes and blend ratios are restricted to
\begin{eqnarray}
0 & \leq & F_\rmn{S}^{(s)} \quad \leq \quad F_\rmn{r}^{(s)}/\zeta_\rmn{f,B}\,, \\
F_\rmn{B}^{(s)} & \geq & F_\rmn{f}^{\star\,(s)} -
F_\rmn{r}^{(s)}\,
g_{\rmn{f},\rmn{min}}^\star
 \quad \geq \quad 0\,,\\
F_\rmn{base}^{(s)} & \geq & F_\rmn{f}^{\star\,(s)}-
F_\rmn{r}^{(s)}\,\left(g_{\rmn{f},\rmn{min}}^\star - \frac{1}{\zeta_\rmn{f,B}}\right)
 \quad \geq \quad 
F_\rmn{S}^{(s)}\,, \\
g^{(s)} & \geq & 
\left(\frac{F_\rmn{f}^{\star\,(s)}}{F_\rmn{r}^{(s)}}-
g_{\rmn{f},\rmn{min}}^\star\right)\,\zeta_\rmn{f,B}
\quad \geq \quad 0\,.
\end{eqnarray}
By definition, 
\begin{equation}
g_{\rmn{f},\rmn{min}}^\star\leq 
\frac{F_\rmn{f}^{\star\,(s)}}{F_\rmn{r}^{(s)}}\,,
\end{equation}
where the equality holds for the lightcurve $s_\rmn{min}$
involving the minimal value 
$g^{\star\,(s_\rmn{min})}_\rmn{f} =
g_{\rmn{f},\rmn{min}}^\star$.

With $\phi(\zeta_\rmn{f})$ given by Eq.~(\ref{eq:phiofzeta}) if one takes
the fold-caustic model parameter ${\hat \omega}_\rmn{f}^\star$ into account,
the lower limit on $\zeta_\rmn{f}$ due to non-negative backgrounds translates
into a limit $\phi(\zeta_\rmn{f,B}) = \phi(\zeta_\rmn{f,B})$ for the
caustic crossing angle $\phi$, which replaces
the boundary at $\Gamma_\rmn{f}$ (for $\Gamma_\rmn{f} \in (0,\upi)$) or
$\Gamma_\rmn{f} - \upi$ (for $\Gamma_\rmn{f} \in (\upi,2\upi)$) as shown
in Table~\ref{tab:restrictanglevarblend}.
With $\zeta_\rmn{f} = |\zeta_\rmn{f,ref}|$ for
$\Gamma_\rmn{f} = 0$ or $\Gamma_\rmn{f} = \upi$ irrespective of the caustic
crossing angle $\phi$, no restriction on $\phi$ results if
$|\zeta_\rmn{f,ref}| \geq \zeta_\rmn{f,B}$, whereas the configuration 
$(d, q, \ell, \zeta_\rmn{f})$ provides no viable solution
otherwise.

\begin{table*}
\caption{Constraints on the caustic crossing angle $\phi$ due to a limit on the variation of
the magnification of non-critical images and the requirement of non-negative background fluxes.}
\label{tab:restrictanglevarblend}
\begin{tabular}{ccccc}
\hline
${\hat \omega}_\rmn{f}^\star >0$ & $\Gamma_\rmn{f} \in (0,\upi)$ &
$0 < \Gamma_\rmn{f} < \phi_\rmn{var} < \upi$ &
$0 < \Gamma_\rmn{f} < \phi_\rmn{B} < \upi$ &
$0 < \Gamma_\rmn{f} < \phi_\rmn{B} \leq \phi < \phi_\rmn{var} < \upi$ \\[0.18ex]
${\hat \omega}_\rmn{f}^\star >0$ & $\Gamma_\rmn{f} \in (\upi,2\upi)$ &
$0 < \phi_\rmn{var} < \Gamma_\rmn{f}-\upi  < \upi$ &
$0 < \phi_\rmn{B} < \Gamma_\rmn{f}-\upi  < \upi$ &
$0 < \phi_\rmn{var} < \phi \leq \phi_\rmn{B} < \Gamma_\rmn{f}-\upi <  \upi$ \\[0.18ex]
${\hat \omega}_\rmn{f}^\star <0$ & $\Gamma_\rmn{f} \in (0,\upi)$ &
$0 < \phi_\rmn{var} < \Gamma_\rmn{f}  < \upi$ &
$0 < \phi_\rmn{B} < \Gamma_\rmn{f}  < \upi$ &
$0 < \phi_\rmn{var} < \phi \leq \phi_\rmn{B} < \Gamma_\rmn{f} <  \upi$ \\[0.18ex]
${\hat \omega}_\rmn{f}^\star <0$ & $\Gamma_\rmn{f} \in (\upi,2\upi)$ &
$0 < \Gamma_\rmn{f} -\upi< \phi_\rmn{var} < \upi$ &
$0 < \Gamma_\rmn{f} -\upi < \phi_\rmn{B} < \upi$ &
$0 < \Gamma_\rmn{f} -\upi< \phi_\rmn{B} \leq \phi < \phi_\rmn{var} < \upi$\\[0.18ex]
\hline
\end{tabular}

\medskip
The upper limit $\zeta_\rmn{f} < \zeta_\rmn{f,var}$ which avoids
 $A^\rmn{p}_\rmn{other}(t) < 1$ over the temporal range covered by observations and
 the lower limit $\zeta_\rmn{f} \geq \zeta_\rmn{f,B}$ which avoids negative background fluxes
 translate to the shown
 limits on the caustic
 crossing angle $\phi_\rmn{var} = \phi(\zeta_\rmn{f})$ or
 $\phi_\rmn{B} = \phi(\zeta_\rmn{f,B})$, respectively, with $\phi(\zeta_\rmn{f})$ given by
 Eq.~(\ref{eq:phiofzeta}),
 where the allowed range depends on the temporal variation of the magnification due to
 the non-critical images
 described by the fold-caustic model parameter $\hat \omega_\rmn{f}^\star$ and the
 the orientation of their gradient $(\vec \nabla A)_\rmn{f}$ given by the angle $\Gamma_\rmn{f}$
 between the caustic inside normal $\vec n_\rmn{f}$ and $(\vec \nabla A)_\rmn{f}$.
 The variation limit $\phi_\rmn{var}$ replaces the boundary at $\phi = 0$ or $\phi = \upi$,
 whereas the blending limit $\phi_\rmn{B}$ replaces the boundary at $\phi = \Gamma_\rmn{f}$
 or $\phi = \Gamma_\rmn{f}-\upi$.
\end{table*}

\subsection{Constraints from data outside the caustic-passage region}
\label{sec:constraintsother}
The tails of the lightcurve of a binary lens microlensing event approach those of a single 
lens event. From data taken in the corresponding regions 
(far outside the caustic passages),
the baseline fluxes $F_\rmn{base}^\rmn{(s)}$ are easily determined 
while data of better quality will allow to determine the time-scale of motion
$t_\rmn{E}$ and the blend ratios
$g^\rmn{(s)}$, yielding measurements of the source fluxes $F_\rmn{S}^{(s)}$ and the
background fluxes $F_\rmn{B}^{(s)}$.
As will be pointed out later,
measurements of these parameters are vital for constraining the mass ratio $q$ and the
angular separation $d\,\theta_\rmn{E}$ of the binary lens and for predicting
other caustic passages.

If the source flux $F_\rmn{S}^{(k)}$ is known for (at least) one site $k$,
the caustic rise parameter $\zeta_\rmn{f}$ is determined as
\begin{equation}
\zeta_\rmn{f} = \frac{F_\rmn{r}^{(k)}}{F_\rmn{S}^{(k)}} \,, \label{eq:zfFS} \\
\end{equation}
while it follows from the background flux $F_\rmn{B}^{(k)}$,
baseline flux $F^{(k)}_\rmn{base}$ 
or blend ratio $g^{(k)}$
with $A_\rmn{f}$ as
\begin{eqnarray}
\zeta_\rmn{f}  & = & \frac{F_\rmn{r}^{(k)}}
{F_\rmn{f}^{\star\,{(k)}} + F_\rmn{r}^{(k)}\,{\hat \omega_\rmn{f}^\star}\,
t_\star^\perp - 
F^{(k)}_\rmn{B}}\,A_\rmn{f}\,, \label{eq:zfFB} \\
\zeta_\rmn{f}  & = & \frac{F_\rmn{r}^{(k)}}{F_\rmn{f}^{\star\,{(k)}}
+ F_\rmn{r}^{(k)}\,{\hat \omega_\rmn{f}^\star}\,
t_\star^\perp - 
F^{(k)}_\rmn{base}}\,\left(A_\rmn{f} - 1\right)\,, \\
\zeta_\rmn{f}  & = & \frac{F_\rmn{r}^{(k)}}{F_\rmn{f}^{\star\,{(k)}}
+ F_\rmn{r}^{(k)}\,{\hat \omega_\rmn{f}^\star}\,
t_\star^\perp} 
\,\left(A_\rmn{f} + g^{(k)}\right)\,.  \label{eq:zfg}
\end{eqnarray}
The conditions $F_\rmn{S}^{(k)} > 0$ and $A_\rmn{f} > 1$ imply
$F_\rmn{fold}(t_\rmn{f}) = F_\rmn{S}^{(k)} A_\rmn{f} + F_\rmn{B}^{(k)}$ 
$= F_\rmn{f}^{\star\,{(k)}}
+ F_\rmn{r}^{(k)}\,{\hat \omega}_\rmn{f}^\star\,
t_\star^\perp > F_\rmn{B}^{(k)}$ and 
$F_\rmn{f}^{\star\,{(k)}}
+ F_\rmn{r}^{(k)}\,{\hat \omega}_\rmn{f}^\star 
t_\star^\perp > F_\rmn{base}^{(k)}$, so that $\zeta_\rmn{f} > 0$ is ensured.
If $F_\rmn{f}^{\star\,{(k)}} + F_\rmn{r}^{(k)}\,{\hat \omega}_\rmn{f}^\star = 0$, 
$\zeta_\rmn{f}$ cannot be determined from $g^{(k)}$, whereas $A_\rmn{f}$
follows directly as  $A_\rmn{f} = -g^{(k)}$ in this case.

The coefficient $A_\rmn{f}^\star = A_\rmn{f}
- \zeta_\rmn{f}\,{\hat \omega}_\rmn{f}^\star\,t_\star^\perp$ in general depends
on the fold-caustic
model parameters ${\hat \omega}_\rmn{f}$ and $t_\star^\perp$ and on the choice of
$\zeta_\rmn{f}$, while for ${\hat \omega}_\rmn{f}^\star = 0$, this
relation reduces to 
$A_\rmn{f}^\star = A_\rmn{f}$ and 
$A_\rmn{f}^\star$  follows directly from the choice of $(d,q,\ell)$.

With the values of the caustic rise fluxes $F_\rmn{r}^{(s)}$,
the determination of $\zeta_\rmn{f}$ fixes all source fluxes $F_\rmn{S}^{(s)}$
according to Eq.~(\ref{eq:FSzf}),
and together with the caustic background fluxes $F_\rmn{f}^{\star\,{(k)}}$ and
$A_\rmn{f}^\star$, all
background fluxes $F_\rmn{B}^{(s)}$, baseline fluxes $F_\rmn{base}^{(s)}$ and
the blend ratios $g^{(s)}$ follow according to Eqs.~(\ref{eq:FBzf}) to ~(\ref{eq:gzf}).
Therefore, a measurement of
one of these quantities for a single lightcurve  fixes all of these
quantities for the full set of lightcurves. 

The fluxes $F_\rmn{r}^{(s)}$ and $F_\rmn{f}^{\star\,(s)}$
obtained from modelling the data near a caustic passage
contain information about the relative source
fluxes and the relative blending between the different lightcurves, while their
absolute values are related to $\zeta_\rmn{f}$ and $A_\rmn{f}^\star$.
Explicitly, Eq.~(\ref{eq:defFr}) implies that the source flux for any lightcurve $s$
is determined by the source flux for a specific lightcurve $k$ with 
$F_\rmn{r}^{(s)}$ and $F_\rmn{r}^{(k)}$ as
\begin{equation}
F^{(s)}_\rmn{S} = 
\frac{F_\rmn{r}^{(s)}}{F_\rmn{r}^{(k)}}\,F^{(k)}_\rmn{S}
\label{eq:FSall}\,,
\end{equation}
while Eqs.~(\ref{eq:flbasesource}) and~(\ref{eq:flblendsource}) yield the background and
source fluxes for any lightcurve from the corresponding values of a specific lightcurve
with $F_\rmn{r}^{(s)}$, $F_\rmn{r}^{(k)}$, 
$F^{\star(s)}_\rmn{f}$, and $F^{\star(k)}_\rmn{f}$ as  
\begin{eqnarray}
F^{(s)}_\rmn{B} & = & F^{\star(s)}_\rmn{f} -
\frac{F_\rmn{r}^{(s)}}{F_\rmn{r}^{(k)}}\,\left(
F^{\star(k)}_\rmn{f} - F^{(k)}_\rmn{B}\right)
\,, \label{eq:FBall} \\
F^{(s)}_\rmn{base} & = & F^{\star(s)}_\rmn{f} -
\frac{F_\rmn{r}^{(s)}}{F_\rmn{r}^{(k)}}\,\left(
F^{\star(k)}_\rmn{f} - F^{(k)}_\rmn{base}\right)
\,. 
\label{eq:fbaseall}
\end{eqnarray}

By means of Eqs.~(\ref{eq:zfFS}) to~(\ref{eq:zfg}), $A_\rmn{f}$ can be determined
if two of the parameters
$F^{(k)}_\rmn{S}$,
$F^{(k)}_\rmn{B}$, $F^{(k)}_\rmn{base}$, or $g^{(k)}$ are known for the same lightcurve,
which is already the case if two of these parameters are known for any pair
of different lightcurves.

Similar to the fluxes, the blend ratios $g^{(s)}$ for all sites can be deduced from the
value for any site $g^{(k)}$ as
\begin{equation}
g^{(s)}  =  
\frac{F_\rmn{r}^{(k)}}{F_\rmn{r}^{(s)}}\,
\frac{F^{\star(s)}_\rmn{f} 
+ F_\rmn{r}^{(s)}\,{\hat \omega}_\rmn{f}^\star\,t_\star^\perp}
{F^{\star(k)}_\rmn{f}
+ F_\rmn{r}^{(k)}\,{\hat \omega}_\rmn{f}^\star\,t_\star^\perp}\,
\left(A_\rmn{f} + g^{(k)}\right) - A_\rmn{f}\,,
\label{eq:gall}
\end{equation}
but in contrast,
this relation involves the parameter $A_\rmn{f}$,
so that the measurement of two different
blend ratios $g^{(k)} \neq g^{(l)}$ yields not only $\zeta_\rmn{f}$ as a function
of $A_\rmn{f}$, but also
\begin{eqnarray}
A_\rmn{f} & = &
\left[F_\rmn{r}^{(l)}\,(F^{\star(k)}_\rmn{f}+
F_\rmn{r}^{(k)}\,{\hat \omega}_\rmn{f}^\star\,t_\star^\perp)\,g^{(l)}\right.\,- \nonumber \\
& & \quad -\,\left.
F_\rmn{r}^{(k)}\,(F^{\star(l)}_\rmn{f}
+ F_\rmn{r}^{(l)}\,{\hat \omega}_\rmn{f}^\star\,t_\star^\perp)\,g^{(k)}\right] \nonumber \,\Big/
\nonumber \\
& & 
\Big/\,
\left[F_\rmn{r}^{(k)}\,
(F^{\star(l)}_\rmn{f} 
+ F_\rmn{r}^{(l)}\,{\hat \omega}_\rmn{f}^\star\,t_\star^\perp)\right.\,- \nonumber \\
& & 
\quad -\,\left. F_\rmn{r}^{(l)}\,
(F^{\star(k)}_\rmn{f}
+F_\rmn{r}^{(k)}\,{\hat \omega}_\rmn{f}^\star\,t_\star^\perp)\right]
\label{eq:Afsfromg}
\end{eqnarray}
itself.
The condition $g^{(k)} \neq g^{(l)}$ ensures that the denominator does not vanish.
For $F^{\star(k)}_\rmn{f}+F_\rmn{r}^{(k)}\,{\hat \omega}_\rmn{f}^\star\,
t_\star^\perp = 0$, Eq.~(\ref{eq:Afsfromg}) reveals
$A_\rmn{f} = - g^{(k)}$, making the measurement of $g^{(l)}$ redundant
for obtaining $A_\rmn{f}$ in this case.

For a given $(d,q,\ell)$,
the determination of $\zeta_\rmn{f}$
not only yields all fluxes and blend ratios,
but also the source size parameter $\rho_\star$ and
the time-scale of perpendicular motion $t_\rmn{E}^\perp$. 
If ${\hat \omega}_\rmn{f}^\star$ is disregarded,
the parameter search
reduces to the four-dimensional subspace 
$(d,q,\ell,\phi)$.
Otherwise, with a measured  
${\hat \omega}_\rmn{f}^\star$, $\zeta_\rmn{f}$ determines $\phi$ with $R_\rmn{f}$,
${\vec n}_\rmn{f}$ and $(\vec \nabla A)_\rmn{f}$, so that 
the proper motion $\mu$, the time $t_\star$ 
in which the source moves by its angular radius, the time-scale of motion $t_\rmn{E}$, and the whole
lightcurve are defined, while the search of parameters is reduced to the 
three-dimensional subspace $(d,q,\ell)$.

A configuration $(d,q,\ell)$ can be discarded if $A_\rmn{f}$ does not match the value
determined from a measurement of 
two fluxes or two different blend ratios 
or from a flux and a blend ratio.

A measurement of the time-scale of motion $t_\rmn{E}$ yields a relation between
the caustic crossing angle $\phi$ and the caustic rise parameter $\zeta_\rmn{f}$,
depending on $R_\rmn{f}$ defined by $(d,q,\ell)$ 
and the fold-caustic model parameter $t_\rmn{f}$, thereby
reducing the number of free parameters by one.

The position of another caustic passage provides a relation between $t_\rmn{E}$ and $\phi$ 
which also implies a relation between $\zeta_\rmn{f}$ and $\phi$ with $R_\rmn{f}$.

Since each of the measurements of ${\hat \omega}_\rmn{f}^\star$, $t_\rmn{E}$, and
the position of another caustic passage provide a relation between 
$\zeta_\rmn{f}$ and $\phi$ depending on $R_\rmn{f}$, ${\vec n}_\rmn{f}$ and
$({\vec \nabla A})_\rmn{f}$,
and a measurement of one of the fluxes $F_\rmn{S}^{(s)}$, $F_\rmn{B}^{(s)}$,
$F_\rmn{base}^{(s)}$ or a blend ratio $g^{(k)}$ yields $\zeta_\rmn{f}$,
both $\zeta_\rmn{f}$ and $\phi$ are determined and
the space of free parameters reduces to the three-dimensional space $(d,q,\ell)$ with two of these
parameters being determined, while the determination of more than two of these parameters
yields constraints for $(d,q,\ell)$ which characterize   
the binary lens
model and the position of the singularity on the caustic. 
With the restriction of non-negative background fluxes 
$F_\rmn{B}^{(s)} \geq 0$ and $\zeta_\rmn{f}$ being determined, 
Eq.~(\ref{eq:zetablendlimit})
yields a limit 
\begin{equation}
A_\rmn{f} < 
\zeta_\rmn{f}\,
(g_{\rmn{f},\rmn{min}}^\star
+{\hat \omega}_\rmn{f}^\star\,t_\star^\perp)\,,
\end{equation}
which puts a restriction on the choice of the binary lens and fold singularity 
parameters $(d,q,\ell)$ for a given caustic crossing angle $\phi$.

The arising constraints for $(d,q,\ell)$ and the power of the prediction of other caustic passages
are further discussed in the following section.

\section{Predictive power}
\label{sec:predpow}

A dense photometric sampling of
binary lens microlensing events that involve fold-caustic passages 
provides opportunities for resolving the stellar atmosphere of the source
star yielding measurements of limb-darkening coefficients, for measuring the 
proper motion $\mu$ of the source relative to the lens, and for obtaining 
the physical properties of the lens system such as the total mass, mass ratio,
semimajor axis, and orbital period. From all lens properties,
only the mass ratio $q$ and the 
separation parameter $d = \delta/\theta_\rmn{E}$ are observable in 
the lightcurve, whereas the lens mass $M$ and the distance $D_\rmn{L}$ are convolved
into the time-scale of motion $t_\rmn{E}$ which however can be disentangled from
an assessment of effects in the lightcurve caused by the parallactic motion.
The measurement of semimajor axis and orbital period is hampered by the fact that
the lightcurve depends on the instantaneous angular separation only 
(except for some small effects
caused by orbital motion) leaving
inclination and  eccentricity of the binary system unconstrained,
so that these quantities can only be assessed statistically.

The determination of the
local properties of the lens mapping near the fold singularity parametrized as
described in 
Sect.~\ref{sec:parametrization} 
allows to study the atmosphere of the observed source star and to obtain a lower limit
on its proper motion $\mu$ relative to the caustic
(i.e.~relative to the lens system approximately), namely
its component $\mu^\perp$ perpendicular to the caustic.

A powerful test of stellar atmosphere models requires a dense coverage of
the lightcurve during the
caustic passage \citep[e.g.][]{Rhie:ld}. 
In order to be able to schedule the necessary observations, including
taking some high-resolution spectra which will provide a deep probe of the
chemical composition of the source star by means of observable variations in 
associated spectral lines, caustic passages need to be predicted some time ahead.
Unless the source trajectory approaches a cusp, fold-caustic passages appear in pairs, 
comprised of a caustic entry and a
subsequent caustic exit. Caustic entries are practically unpredictable before
they actually occur, so
that their beginning phase can only be caught by chance and immediate action has to be taken to 
follow the lightcurve over the caustic entry.
Since the behaviour of the lightcurve around a caustic passage depends on local properties
only, the data taken during the caustic entry contains practically no information about the following
caustic exit \citep[c.f.][]{PLANET:sol,JM:2ndcc}. 
However, caustic exits can be predicted 
using the data on the rise to the caustic passage peak. Once such a rise has progressed,
the parameters $t_\rmn{f}^\star = t_\rmn{f}$, $F_\rmn{r}^{(s)}$, and
$F_\rmn{f}^{\star\,(s)}$ can be determined for a point source ($t_\star^\perp = 0$) and
${\hat \omega}_\rmn{f}^\star = 0$. When the source reaches the region where the 
curvature of the lightcurve changes sign, it becomes possible to assess the source size by
including $t_\star^\perp > 0$ in the fit, thereby obtaining a prediction for the end of the
caustic exit at $t_\rmn{f}^\star$ which however depends on the amount of limb darkening 
which might be determined from a fit to the data
as well as ${\hat \omega}_\rmn{f}^\star$ at later stages.

The data taken near the caustic passage also neither provide constraints on the 
mass ratio $q$ and the 
angular separation parameter $d = \delta/\theta_\rmn{E}$
which characterize the binary lens nor on the location of the fold singularity ${\vec y}_\rmn{f}$ on
the corresponding caustic described by the parameter $\ell$. The parameters $(d,q,\ell)$ are related
to the fold-caustic model parameters by means of the local properties $R_\rmn{f}$, 
${\vec n}_\rmn{f}$, $A_\rmn{f}$
and $({\vec \nabla} A)_\rmn{f}$ which are not observables themselves but affect the lightcurve
through convolutions with the source size parameter $\rho_\star$ and the source and background fluxes 
$F_\rmn{S}^{(s)}$ and $F_\rmn{B}^{(s)}$. 
The insensitivity to source motion parallel to the caustic
leaves the caustic crossing angle $\phi$ undetermined.
The parameters $(d,q,\ell,\phi)$ fix the binary lens system, the spatial
position of the source trajectory, and with 
$t_\rmn{f} = t_\rmn{f}^\star \pm t_\star^\perp$ 
the point of time where
the source center crosses. 
The unknown source flux (or baseline flux)
 results in a free caustic rise parameter $\zeta_\rmn{f}$.
With the rise fluxes $F_\rmn{r}^{(s)}$, $\zeta_\rmn{f}$ yields the source fluxes
$F_\rmn{S}^{(s)}$, and together 
with the caustic baseline fluxes $F_\rmn{f}^{\star\,(s)}$ and
$A_\rmn{f}^\star$,\footnote{$A_\rmn{f}^\star = A_\rmn{f}-\zeta_\rmn{f}\,
{\hat \omega}_\rmn{f}^\star\,t_\star^\perp$, which reduces to $A_\rmn{f}^\star = A_\rmn{f}$
for ${\hat \omega}_\rmn{f} = 0$.}
it yields the background fluxes $F_\rmn{B}^{(s)}$. Together with
$R_\rmn{f}$, $\zeta_\rmn{f}$ yields the source size parameter
$\rho_\star = \theta_\star 
/\theta_\rmn{E}$ and the time-scale of transverse motion $t_\rmn{E}^\perp$, while 
the time-scale of motion $t_\rmn{E}$ also depends on $\phi$.
 
Since the observed caustic-passage time  
$t_\star^\perp$
is the product of the transverse time-scale of motion $t_\rmn{E}^\perp = t_\rmn{E}/(\sin \phi)$
and the source size
parameter $\rho_\star$,
smaller sources with smaller transverse proper motion
or larger sources with larger transverse proper motion can yield the same lightcurve.
Moreover, the insensitivity to the crossing angle means that the same transverse proper motion can be 
achieved for any angle by a corresponding choice of the full proper motion.

While the 
determination of the model parameters for the lightcurve near the caustic passage alone cannot 
yield any predictions for its remaining parts, the combination   
with a few simple characteristics of the data outside the caustic-passage region  
however provides some vital constraints.

Restrictions of the parameters $(d,q,\ell)$ can arise from independent measurements of (combinations
of) the
local properties $R_\rmn{f}$, ${\vec n}_\rmn{f}$, $A_\rmn{f}$
or $({\vec \nabla} A)_\rmn{f}$. As pointed out in Sect.~\ref{sec:constraintsother},
measurements of one baseline flux $F_\rmn{base}^{(k)}$
and the blend ratio $g^{(k)}$ or of two blend ratios $g^{(k)}$ and $g^{(l)}$
yield $A_\rmn{f}$,\footnote{$F_\rmn{base}^{(k)}$ is 
easily obtained as observable of the lightcurve, while $g^{(k)}$ is also determined with
$F_\rmn{base}^{(k)}$ and a measurement of either the source flux $F_\rmn{S}^{(k)}$ or
the background flux $F_\rmn{B}^{(k)}$.}
while measurements of ${\hat \omega}_\rmn{f}^\star$,
one baseline flux $F_\rmn{base}^{(k)}$, and the time-scale of motion $t_\rmn{E}$
provide a relation between $R_\rmn{f}$, $A_\rmn{f}$
${\vec n}_\rmn{f}$, and $({\vec \nabla} A)_\rmn{f}$.

For fixed fold-caustic parameters ($t_\star^\perp$, $F_\rmn{r}^{(s)}$,
$F_\rmn{f}^{\star\,(s)}$, ${\hat \omega}_\rmn{f}^\star$), 
the full lightcurve is defined with the 
choice of $(d,q,\ell,\zeta_\rmn{f},\phi)$,
so that all of its features such as the position, duration and
peak flux of other caustic passages or other types of peaks, and the
time-scale of motion $t_\rmn{E}$ are determined.

By requiring a match of the model lightcurve with the observed data,
the determination of $\zeta_\rmn{f}$ and $\phi$ or the provision of a relation between
these parameters together with limits, arising from the appropriate sign of
${\dot A}_\rmn{f}^\star$ (that allows to fulfill $\zeta_\rmn{f}> 0$ with the sign
of ${\hat \omega}_\rmn{f}^\star$) 
or the restriction to positive background fluxes, therefore yields restrictions 
for the choice of $(d,q,\ell)$.
Alternatively, the restrictions on $(d,q,\ell,\zeta_\rmn{f},\phi)$ constrain
the behaviour of the lightcurve outside the caustic-passage region, so that
in particular limits on the position, duration, and peak flux of other
caustic passages arise. Due to intrinsic ambiguities for binary lens models
\citep{Do:Ambig,Do99:CR}, 
a unique solution however
cannot be expected and different scenarios will be possible. 

A relation between $\zeta_\rmn{f}$ and $\phi$ is provided
by a measurement of ${\hat \omega}_\rmn{f}^\star$ (which also yields
an upper limit on $t_\rmn{E}$) or of $t_\rmn{E}$, and
together with a measurement of a baseline flux $F_\rmn{base}^{(k)}$, 
$\zeta_\rmn{f}$ and $\phi$ are determined for given $(d,q,\ell)$.

Determining ${\hat \omega}_\rmn{f}^\star$ and $t_\rmn{E}$ fixes
$\zeta_\rmn{f}$ and $\phi$ with $R_\rmn{f}$,
${\vec n}_\rmn{f}$, and $({\vec \nabla} A)_\rmn{f}$.
If a baseline flux $F_\rmn{base}^{(k)}$ is known in addition,
a relation between $R_\rmn{f}$, ${\vec n}_\rmn{f}$, $A_\rmn{f}$,
and $({\vec \nabla} A)_\rmn{f}$ results.

The observed position of another caustic passage provides a relation between
$t_\rmn{E}$ and $\phi$ which implies a relation between $\zeta_\rmn{f}$ and $\phi$
with $R_\rmn{f}$. Combining this information with measurements of one of the parameters
${\hat \omega}_\rmn{f}^\star$, $t_\rmn{E}$ or a baseline flux
$F_\rmn{base}^{(k)}$ fixes $\zeta_\rmn{f}$ and $\phi$, while a combination with
more than one of these parameters also yields relations between 
$R_\rmn{f}$, ${\vec n}_\rmn{f}$, $A_\rmn{f}$,
and $({\vec \nabla} A)_\rmn{f}$, thereby restricting $(d,q,\ell)$.
Other features of the additional caustic passage such as its duration, its peak flux, or the
flux and rate of change of flux at the limb of the source entering or exiting the caustic, 
yield additional restrictions which can become quite powerful
\citep[c.f.~][]{PLANET:EB5mass}.

If $\zeta_\rmn{f}$ is determined, the requirement of non-negative background flux
$F_\rmn{B}^{(s)} \geq 0$ constrains $A_\rmn{f}$, so that
the choice of $(d,q,\ell)$ is restricted if $\phi$ is determined in addition.

For a worked example, 
some considerations about the power of different characteristics of the photometric
data outside the caustic-passage region for determining the lens properties and
predicting other caustic passages have been made
by \citet{PLANET:sol}.

\section{Summary}
\label{sec:summary}

Dense high-precision photometric sampling of microlensing lightcurves during fold-caustic
passages can provide both a measurement of the proper motion $\mu$ between lens and source star
and a probe of the stellar atmosphere of the source star, e.g.~parametrized by limb-darkening
coefficients, allowing to test existing theoretical models. Making use of the 
alerts supplied by the current microlensing surveys OGLE and MOA, 
a follow-up campaign like PLANET is capable of providing 10--15 such measurements 
per year on different types of galactic stars.

While the duration of the caustic passage $t_\star^\perp$ 
obtained from modelling the data in the caustic-passage
region and of the angular size of the source $\theta_\star$, which can be obtained by spectral
typing, directly yield the proper motion $\mu^\perp = \theta_\star/t_\star^\perp$ 
of the source perpendicular to the caustic,
the full proper motion $\mu = \mu^\perp/(\sin \phi)$ only follows with a determination of the
caustic crossing angle $\phi$ from a model of the full lightcurve.

During fold-caustic passages, the microlensing lightcurve shows an
increased sensitivity to the parallactic motion of the Earth around the Sun and 
to the orbital motion of the binary lens whose measurement
breaks the degeneracy that remains between lens mass $M$ and distance $D_\rmn{L}$
after determination of the proper motion $\mu$ \citep{Do:Rotate,PLANET:EB5mass}.
Therefore, microlensing events that involve caustic passages are prime objects for
determining the 
mass function and the phase-space distribution of stellar populations whose
constituents act as
gravitational microlenses.

Moreover, the prominent feature of caustic-passage peaks allows 
a proper determination of binary lens 
parameters and with the possibility to measure $\mu$, $M$ and $D_\rmn{L}$ 
individually, a reasonable opportunity is provided for obtaining
statistical distributions of properties of stellar and sub-stellar binaries such as
their mass, mass ratio, semimajor axis, and orbital period.

For a given distribution of lens masses, the lens mapping maps the position of an observed
image of a source star to its true position. While there is a unique true source position for
any observed image, a source may have several images. While the lens mapping is regular
(i.e.~its Jacobian determinant does not vanish), it can be locally inverted, so that the 
number of images only changes if the source reaches a singular point of the lens mapping.
The lowest-order singularities are folds which form closed curves called critical curves 
in the space of image positions and caustics in the space of true source positions.
When a point source crosses a fold caustic from the 'inside' to the 'outside', the
magnification of two of its images 
diverges with the inverse square-root of the distance perpendicular to the caustic, where
these two images become critical, while they disappear just as the source exits the caustic.
The fold caustic is characterized by the single parameter $R_\rmn{f}$ 
describing its strength, i.e.~a proportionality factor between the magnification and the
inverse square-root of the perpendicular source-caustic distance.

The magnification of the critical images of an extended source with source size parameter 
$\rho_\star = \theta_\star/\theta_\rmn{E}$ is characterized by a universal caustic profile 
function describing the shape of the lightcurve during a caustic passage which is 
characteristic for any given radial brightness profile of the source, while
for the same 
caustic strength $R_\rmn{f}$, the magnification is proportional to $\rho_\star^{-1/2}$.
If the luminosity of the source at the limb does not vanish, a discontinuous change in the
slope of the lightcurve is produced as the leading limb of the caustic hits the caustic.
On the source further entering the caustic, the magnification rises to a peak and then drops 
asympotically proportional to the inverse square-root of the perpendicular
distance of the source center from the caustic.
 
The caustic-passage regions of $n$ observed 
lightcurves (for different sites and/or spectral bands)
during a caustic passage can be characterized by the $3+2n$ parameters 
($t_\rmn{f}^\star$, $t_\star^\perp$,  $F_\rmn{r}^\rmn{(s)}$,
$F_\rmn{f}^{\star\,\rmn{(s)}}$, ${\hat \omega}^\star_\rmn{f}$).
The source crosses the caustic during the timespan $2\,t_\star^\perp$,
where its leading limb enters or its trailing limb exits
the caustic at time $t_\rmn{f}^\star$ where the flux is given by
$F_\rmn{f}^{\star\,\rmn{(s)}}$.
$F_\rmn{r}^\rmn{(s)}$ denotes
the flux of the two critical images at a unit time ${\hat t}$
after the source center enters or before it exits the caustic
that are produced 
if the extended source is replaced
by a point source at its center, and
the parameter ${\hat \omega}^\star_\rmn{f}$
measures the rate of the (constant) change of flux in units of $F_\rmn{r}^{(s)}$ 
due to the non-critical images for the source moving inwards, while this rate
is given by $-{\hat \omega}^\star_\rmn{f}$ for a source moving outwards.

Since the influence of the lens mapping on the
microlensing lightcurve near a fold-caustic can be approximated by means of
local properties, the study of stellar 
atmospheres, and in particular the determination of limb-darkening coefficients, does not
require the assessment of a complete set of model parameters that characterize the full light
curve and is not influenced by ambiguities and
degeneracies that are likely to occur in these parameters \citep{DoHi2,Do:Ambig,Do99:CR,PLANET:sol}.
However, this also means that the data taken in the caustic-passage region does not provide
information about the behaviour of the lightcurve outside this region. In particular, 
the caustic-passage data do not allow to obtain predictions about subsequent other caustic 
passages.

For a binary lens model and a fold singularity on its caustic,
all local properties of the lens mapping can be obtained analytically 
by means of derivatives of the Fermat potential. However, these properties 
are not observables themselves but affect the lightcurve
through convolutions with the source size parameter
$\rho_\star$ and the source and background fluxes $F_\rmn{S}^{(s)}$ and $F_\rmn{B}^{(s)}$
which in general are unkown themselves.
The data in the caustic-passage region therefore neither constrains
the mass ratio $q$ of the binary lens and the angular separation $\delta = d\,\theta_\rmn{E}$
between its components, nor the 
location of the fold singularity on the corresponding caustic.

Besides parameters that characterize the source brightness profile, a microlensing event involving
an extended source and a binary lens
is characterized by $7+2n$ parameters.
By modelling the data around an observed fold-caustic passage, thereby fixing
$3+2n$ parameters, or $2+2n$ parameters if ${\hat \omega}^\star_\rmn{f}$ is disregarded,
the search for model parameters describing the full lightcurve
reduces to a search in a four- or five-dimensional subspace
holding the remaining parameters which can be chosen
as $(d,q,\ell,\phi)$ or $(d,q,\ell,\phi,\zeta_\rmn{f})$, respectively,
where $\ell$ characterizes the position of the cusp singularity on the caustic and $\zeta_\rmn{f}$
characterizes the rise of the magnication on the approach to the caustic.
The free choice of $\phi$ reflects the insensitivity to source motion parallel to the caustic,
while the free choice of $\zeta_\rmn{f}$ reflects the freedom in choosing the
source flux.
As demonstrated by \citet{PLANET:sol}, this approach provides an efficient method for obtaining
all suitable models that are consistent with the data of the observed microlensing event.

The determination of a few simple characteristics of the data outside the caustic-passage
regions such as source, background, or baseline fluxes, the blend ratio, the
time-scale of motion $t_\rmn{E}$, or constraints from other characteristic observable features
of the lightcurve such as the position, duration, or peak flux related to other caustic passages
or other peaks yields further restrictions on the parameter space to be investigated and
leads to a reduction of its dimension. 
This information is crucial for obtaining constraints on the mass ratio
$q$ and the 
angular separation parameter $d = \delta/\theta_\rmn{E}$
which characterize the binary lens, and on the properties of subsequent caustic passages.

\section*{Acknowledgements}
For being able to cope with the challenges arising from a long-lasting illness over more than
three years, it was good to know that there are people
around whose motivation in collaborating with me
exceeded that of seeking the own advantage and who would sorely miss me.
I would therefore like to thank those who helped me keeping well-motivated, and also for the
received support on job or grant applications.
In order to avoid weighting individual contributions, no names are listed here, everyone himself 
should be aware of his own personal involvement. 
Even the smallest support had a significant impact on me.
I was really struck by the sudden death of one of my supporters and promise to take care that the
legacy of our interaction will bear fruit.

\begin{table*}
\caption{Used symbols}
\label{tab:overview}
\begin{tabular}{llll}
\hline
& Symbol & Meaning & Definition or prime relation \\[0.18ex] \hline
\multicolumn{4}{l}{\em Caustic-passage lightcurve} \\[0.18ex]
& $t_\rmn{f}^\star$ & time of beginning (end) of caustic entry (exit) &  \\[0.18ex]
& $t_\star^\perp$ & caustic passage half-duration & $t_\star^\perp = \rho_\star t_\rmn{E}^\perp$ 
  \\[0.18ex]
& $\hat \omega_\rmn{f}^\star$ & temporal variation of non-critical images & 
  $\hat \omega_\rmn{f}^\star = \pm {\dot A}_\rmn{f}^\star/\zeta_\rmn{f}$  \\[0.18ex]
& $F_\rmn{r}^{(s)}$ & caustic rise flux & $F_\rmn{r}^{(s)} = F_\rmn{S}^{(s)} \zeta_\rmn{f}$ \\[0.18ex]
& $F^{\star\,(s)}_\rmn{f}$ & flux at time $t_\rmn{f}^\star$  & 
 $F^{\star\,(s)}_\rmn{f} = F_\rmn{fold}^{(s)}(t_\rmn{f}^\star) = 
    F_\rmn{S}^{(s)} A_\rmn{f}^\star + F_\rmn{B}^{(s)}$
 \\[0.18ex]
& $g_\rmn{f}^\star$ & caustic rise flux ratio & $g_\rmn{f}^\star = F^{\star\,(s)}_\rmn{f}/
  F_\rmn{r}^{(s)}$  \\[0.18ex]
& $m^{\star\,(s)}_\rmn{f}$ & magnitude at time  $t_\rmn{f}^\star$  & $m^{\star\,(s)}_\rmn{f} 
= m_\rmn{fold}^{(s)}(t_\rmn{f}^\star)$  \\[0.18ex]
& $\Gamma_{\{p\}}^{(s)}$ & limb-darkening coefficient for profile $\propto \cos^{p} \vartheta$ &  \\[0.18ex]
\multicolumn{4}{l}{\em Caustic-passage auxiliary quantities}  \\[0.18ex]
& $t_\rmn{f}$ & time when source center passes caustic & 
 $t_\rmn{f} = t_\rmn{f}^\star \pm t_\star^\perp$  \\[0.18ex]
& $\zeta_\rmn{f}$ & caustic rise parameter & $\zeta_\rmn{f} = (t_\rmn{r}/{\hat t}\,)^{1/2}$  \\[0.18ex]
& $t_\rmn{r}$ & caustic rise time & $t_\rmn{r} = R_\rmn{f} t_\rmn{E}^\perp$  \\[0.18ex]
\multicolumn{4}{l}{\em Binary-lens lightcurve}  \\[0.18ex]
& $t_\rmn{E}$ & event time-scale & $t_\rmn{E} = \theta_\rmn{E}/\mu$  \\[0.18ex]
& $u_0$ & lens-source impact parameter & $u_0 = \theta_0/\theta_\rmn{E}$  \\[0.18ex]
& $t_0$ & time of closest lens-source approach &   \\[0.18ex]
& $\alpha$ & angle between binary lens axis and source trajectory &  \\[0.18ex]
& $d$ & lens separation parameter & $d = \delta/\theta_\rmn{E}$  \\[0.18ex]
& $q$ & binary lens mass ratio & $q = M_1/M_2$  \\[0.18ex]
& $\rho_\star$ & source size parameter  & $\rho_\star = \theta_\star/\theta_\rmn{E}$  \\[0.18ex]
& $F_\rmn{S}^{(s)}$ & source flux &  \\[0.18ex]
& $F_\rmn{B}^{(s)}$ & background flux &  \\[0.18ex]
& $F_\rmn{base}^{(s)}$ & baseline flux & 
 $F_\rmn{base}^{(s)} = F_\rmn{S}^{(s)} + F_\rmn{B}^{(s)}$  \\[0.18ex]
& $g^{(s)}$ & blend ratio & $g^{(s)} = F_\rmn{B}^{(s)}/F_\rmn{S}^{(s)}$  \\[0.18ex]
\multicolumn{4}{l}{\em Local lens properties}  \\[0.18ex]
& $R_\rmn{f}$ & caustic strength &  \\[0.18ex]
& $\vec n_\rmn{f}$ & caustic inside normal  \\[0.18ex]
& $\vec y_\rmn{f}$ & location of source singularity & $\vec y_\rmn{f} = \vec y(t_\rmn{f})$  \\[0.18ex]
& $A_\rmn{f}$ & magnification due to non-critical images at $\vec y_\rmn{f}$ or $t_\rmn{f}$ 
& $A_\rmn{f} = A^\rmn{p}_\rmn{other}(\vec y_\rmn{f}) = 
A^\rmn{p}_\rmn{other}(t_\rmn{f})$  \\[0.18ex]
& $A_\rmn{f}^\star$ & magnification due to non-critical images at $t_\rmn{f}^\star$
  & $A_\rmn{f}^\star = A^\rmn{p}_\rmn{other}(t_\rmn{f}) \simeq 
    A_\rmn{f} \mp {\dot A}_\rmn{f}^\star t_\star^\perp$
 \\[0.18ex]
& ${\dot A}_\rmn{f}^\star$ & temporal derivative of $A^\rmn{p}_\rmn{other}$ at $t_\rmn{f}^\star$
  & ${\dot A}_\rmn{f}^\star = {\dot A}^\rmn{p}_\rmn{other}(t_\rmn{f}^\star) \simeq  
     {\dot A}^\rmn{p}_\rmn{other}(t_\rmn{f})$ \\[0.18ex]
& $(\vec \nabla A)_\rmn{f}$ & spatial derivative of $A^\rmn{p}_\rmn{other}$ at $\vec y_\rmn{f}$ 
& $(\vec \nabla A)_\rmn{f} = \vec \nabla A^\rmn{p}_\rmn{other}(\vec y_\rmn{f})$  \\[0.18ex]
& $\Gamma_\rmn{f}$ & angle from $\vec n_\rmn{f}$ to $(\vec \nabla A)_\rmn{f}$ &
 $\Gamma_\rmn{f} = \angle(\vec n_\rmn{f},(\vec \nabla A)_\rmn{f})$  \\[0.18ex]
\multicolumn{4}{l}{\em Lens-source motion}  \\[0.18ex]
& $\phi$ & caustic crossing angle \\[0.18ex] 
& $\mu$ & relative lens-source proper motion \\[0.18ex]
& $\mu^\perp$ & lens-source proper motion perpendicular to caustic &
  $\mu^\perp = \mu \sin\phi$  \\[0.18ex]
& $t_\rmn{E}^\perp$ & event time-scale referring to perpendicular motion &
  $t_\rmn{E}^\perp = t_\rmn{E}/(\sin \phi)$  \\[0.18ex]
& $t_\star$ & time for the source to move by its angular radius &
  $t_\star = \rho_\star t_\rmn{E} = \theta_\star/\mu = t_\star^\perp \sin \phi$  \\[0.18ex]
& ${\dot{\vec y}}_\rmn{f}$ & source proper motion vector per $\theta_\rmn{E}$ at $t_\rmn{f}$ &
  ${\dot{\vec y}}_\rmn{f} = \dot{\vec y}(t_\rmn{f})$, $|{\dot{\vec y}}_\rmn{f}| = 1/t_\rmn{E}$  \\[0.18ex]
& $\Gamma_\rmn{t}$ & angle from $\pm \dot{\vec{y}}_\rmn{f}$ to $(\vec \nabla A)_\rmn{f}$ & 
$\Gamma_\rmn{t} = \angle(\pm \dot{\vec{y}}_\rmn{f},(\vec \nabla A)_\rmn{f})$ \\[0.18ex]
\multicolumn{4}{l}{\em Other quantities}  \\[0.18ex]
& $\hat t$ & arbitrary unit time &  \\[0.18ex]
& $M$ & total lens mass & $M = \sum M_i$  \\[0.18ex]
& $M_i$ & individual lens masses  \\[0.18ex]
& $D_\rmn{L}$ & lens distance &  \\[0.18ex]
& $D_\rmn{S}$ & source distance &  \\[0.18ex]
& $\theta_\star$ & angular source radius &  \\[0.18ex]
& $\delta$ & angular separation betwen lens objects &  \\[0.18ex]
& $\theta_0$ & smallest angular lens-source separation &  \\[0.18ex]
& $\theta_\rmn{E}$ & angular Einstein radius & $ \theta_\rmn{E} = 
[(4GM/c^2)(D_\rmn{S}-D_\rmn{L})/(D_\rmn{S} D_\rmn{L})]^{1/2}$
\\ \hline
\end{tabular}

\medskip
Upper (lower) signs correspond to a caustic entry (exit),
$F_\rmn{fold}^{(s)}(t)$ denotes the flux, $m_\rmn{fold}^{(s)}(t)$ the magnitude, 
$A^\rmn{p}_\rmn{other}(t)$ the magnification due to non-critical images,
and $\vec y(t) = \vec \theta(t)/\theta_\rmn{E}$ the source position as a function of time, 
where $\vec \theta(t)$ is the angular position of the source. Among the superscripts and subscripts, 
$\sq_\rmn{f}$ denotes a relation to the fold singularity, 
$\sq^\star$ denotes a reference to the stellar limb, 
$\sq_\star$ denotes a relation to stellar radius, 
$\sq^\perp$ denotes the vector component perpendicular to the fold caustic, and
$\sq^{(s)}$ refers to a specific site and filter.
\end{table*}

\bibliographystyle{mn2e}
\bibliography{mybib}

\end{document}